%% file: modal.tex
\documentclass[10pt]{article}
\usepackage{color}
\usepackage{graphicx}
\usepackage{listings}
\usepackage{xcolor}
\usepackage{tikz}
\usepackage{subcaption}
\usepackage{amsmath}
\usepackage{fixltx2e}
\usepackage{txfonts}
\usepackage{url}
\usepackage{cite}

\usepackage{geometry}
\usetikzlibrary{arrows}
\usetikzlibrary{shapes}

\colorlet{thread0}{green}
\colorlet{thread1}{violet}
\colorlet{thread2}{cyan}
\colorlet{thread3}{orange}

\lstdefinestyle{prettyc}{
  basicstyle=\tiny,
  belowcaptionskip=1\baselineskip,
  breaklines=true,
  frame=L,
  xleftmargin=\parindent,
  language=C,
  showstringspaces=false,
  keywordstyle=\bfseries\color{green!40!black},
  commentstyle=\itshape\color{purple!40!black},
  identifierstyle=\color{blue},
  stringstyle=\color{orange},
  tabsize=4,
  float,
}
\lstdefinestyle{customc}{
  basicstyle=\color{gray}\tiny,
  belowcaptionskip=1\baselineskip,
  breaklines=true,
  frame=L,
  xleftmargin=\parindent,
  language=C,
  showstringspaces=false,
  keywordstyle=\bfseries,
  commentstyle=\itshape,
  tabsize=4,
  moredelim=**[is][\color{blue}]{@}{@},
}

\newcommand{\Intel}{Intel$\textsuperscript{\textregistered}$ }
\newcommand{\Xeon}{Intel Xeon }
\newcommand{\XeonR}{\Intel Xeon$\textsuperscript{\textregistered}$ }
\newcommand{\XeonPhi}{Intel Xeon Phi }
\newcommand{\XeonPhiTM}{\Intel Xeon Phi$\textsuperscript{\texttrademark}$ }
\newcommand{\VTune}{Intel VTune Amplifier XE }
\newcommand{\VTuneTM}{\Intel VTune$\textsuperscript{\texttrademark}$ Amplifier XE }
\newcommand{\x}{$\times\mbox{ }$}

\newcommand{\eq}{\begin{align}}
\newcommand{\qe}{\end{align}}

\newcommand*\swallow[1]{}

\def\a{\alpha}
\def\h{\theta}

\def\px{\approx}

\def\({\left(}
\def\){\right)}
\def\[{\left[}
\def\]{\right]}
\def\<{\left\langle}
\def\>{\right\rangle}

\def\un{{\bf \hat{n}}}

\def\eq{\begin{align}}
\def\qe{\end{align}}

\def\and{\quad \mbox{and} \quad}

\def\bfnl{\kern2pt\overline{\kern-2ptf}_\textrm{NL}}

\def\klist{k_1,k_2,k_3}

\def\barQ{\kern2pt\overline{\kern-2pt{Q}}}

\def\barR{\kern2pt\overline{\kern-2pt{R}}}

\definecolor{dblue}{rgb}{0.2,0,0.5}
\definecolor{dgreen}{rgb}{0.2,0.5,0}

\title{Separable projection integrals for higher-order correlators of the cosmic microwave sky: Acceleration by factors exceeding 100}

\author{
J.P. Briggs$^1$, S.J. Pennycook$^2$, J.R. Fergusson$^1$, J. J\"{a}ykk\"{a}$^1$, E.P.S Shellard$^1$\\
$^1$Department of Applied Mathematics and Theoretical Physics, \\
University of Cambridge, UK\\
$^2$Intel Corporation, UK
}
\begin{document}

\maketitle

\begin{abstract}
\input{abstract}
\end{abstract}

\input{introduction}
\input{background}
\input{implementation}
\input{results}
\input{conclusions}
\input{acknowledgements}
\appendix
\input{appendix}

\bibliographystyle{unsrt}
\bibliography{modal}

\end{document}

%% file: abstract.tex
\noindent We present a case study describing efforts to optimise and modernise ``Modal'', the simulation and analysis pipeline used by the Planck satellite experiment for constraining general non-Gaussian models of the early universe via the bispectrum (or three-point correlator) of the cosmic microwave background radiation. We focus on one particular element of the code: the projection of bispectra from the end of inflation to the spherical shell at decoupling, which defines the CMB we observe today.  This code involves a three-dimensional inner product between two functions, one of which requires an integral, on a non-rectangular domain containing a sparse grid.  We show that by employing separable methods this calculation can be reduced to a one-dimensional summation plus two integrations, reducing the overall dimensionality from four to three. The introduction of separable functions also solves the issue of the non-rectangular sparse grid. This separable method can become unstable in certain scenarios and so the slower non-separable integral must be calculated instead. We present a discussion of the optimisation of both approaches. 

We demonstrate significant speed-ups of $\approx$100\x, arising from a combination of algorithmic improvements and architecture-aware optimisations targeted at improving thread and vectorisation behaviour.  The resulting MPI/OpenMP hybrid code is capable of executing on clusters containing processors and/or coprocessors, with strong-scaling efficiency of 98.6\% on up to 16 nodes.  We find that a single coprocessor outperforms two processor sockets by a factor of 1.3\x  and that running the same code across a combination of both microarchitectures improves performance-per-node by a factor of 3.38\x. By making bispectrum calculations competitive with those for the power spectrum (or two-point correlator) we are now able to consider joint analysis for cosmological science exploitation of new data.

%% file: introduction.tex
\section{Introduction} \label{sec:introduction}

\noindent The current best explanation for the origin of our universe is the inflationary big bang scenario, where it is believed that a period of exponential expansion created the large flat empty universe we see today.  In addition, this model predicts that quantum fluctuations in the energy during this time will be stretched to galactic scales forming the seeds from which all structure grew, from planets and stars through to super-clusters of galaxies.  The statistics of these fluctuations give a window onto the dynamics at play during the birth of our universe. In particular, any deviation of these fluctuations from a Gaussian (\textit{i.e.} Normal) distribution would be direct evidence of interesting new physics.

There has been enormous effort within the community to measure any possible deviations from Gaussianity with the bispectrum, the Fourier transform of the three point correlator, being the favoured statistic. The primary obstacle to na\"{i}ve estimation of the bispectrum is that for the CMB it is 5 dimensional\footnote{this is because it is the average of three vector quantities on a 2D surface (the CMB) which gives you 6 dimensions, enforcing momentum conservation then removes one of these.} and would require $\mathcal{O}(10^{22})$ floating point operations to calculate, which is challenging for the world's largest supercomputers. This can be overcome by using separable approximations for the bispectra, however the projection of the primordial bispectra forward to the time of observation remains a major obstacle to measurement.  There are a multitude of approaches to this which divide into two main categories: ones that require non-Gaussian simulations to train estimators, \cite{0111284, 98083987, 0211399, 08024020, 09063232, 12021478}; and those that use specific simple primordial templates for which projection is tractable, \cite{0305189, 0302223, 09111642, 14052550}.

The Modal method~\cite{fergusson2007, fergusson2010} developed at the University of Cambridge, which is the focus of this paper, and used by the Planck satellite experiment~\cite{PlanckNG2013,PlanckNG2015} is the only general method for constraining these non-Gaussianities from the available data. Its main strength is that by using a general mode expansion it can reduce the evolution from primordial to late times into a matrix projection, allowing us to constrain thousands of theoretical predictions simultaneously.  By using an appropriate basis tuned for the theoretical models of interest, the Modal team have created a fast and efficient way to probe cosmological data for hints of new physics in the early universe.

This paper investigates the optimisation and modernisation of Modal, as part of an effort to accelerate it using \XeonPhiTM coprocessors.  The existing MPI-level parallelism in the original code is not sufficient to enable efficient utilisation of this hardware, and we show that moving to a hybrid MPI/OpenMP implementation can significantly improve performance. For portability reasons, we avoid making any significant code changes that would benefit only one particular hardware platform, and thus the high-level programming languages and techniques that we use apply equally well to \XeonR processors. When compared to the original code on 2\x processor sockets, our code optimisation efforts deliver speed-ups of 1765\x on a single coprocessor and 833\x on 2\x processor sockets in the 2D case; and 108\x on a coprocessor and 83.9\x on 2\x processor sockets in the 3D case. These speed-ups are large enough to significantly impact the rate of scientific discovery at COSMOS, and enable liberal use of the Modal calculation as part of future Monte Carlo pipelines -- something that had previously been considered infeasible.

A number of previous studies have investigated the use of \XeonPhi coprocessors to accelerate other scientific codes~\cite{Pennycook-MD, Patwaray-PARDICLE, Misra-GENOME, Joo-QCD, Apra-nwchem}, and there are many similarities between the optimisations we discuss here and those explored in other domains.  However, we note that many of these studies were performed before the standardisation of OpenMP 4.0 (and thus often rely on manual vectorisation via architecture-specific intrinsics).  This is also the first paper (to the best of our knowledge) to explore the use of \XeonPhi coprocessors for this specific application domain.

The rest of this paper is organized as follows: Section~\ref{sec:background} provides an introduction to the two Modal routines which are being optimised -- the full three dimensional calculation on the sparse non-rectangular domain, and the fast two dimensional version on a dense rectangular domain -- and also provides a high-level introduction to \XeonPhi coprocessors; Section~\ref{sec:implementation} details the optimisation and modernisation of Modal; Section~\ref{sec:results} presents a detailed performance study of the final application, demonstrating its scalability within a node and across multiple nodes; and finally, Section~\ref{sec:conclusions} concludes the paper, and discusses potential new uses for the accelerated version of the code.

%% file: background.tex
\section{Background}\label{sec:background}

\subsection{Direct Integration}

The primary concern of this paper is the efficient calculation of a three-dimensional inner product which concerns the projection of a set of primordial basis bispectra defined at the end of inflation into a set of basis bispectra on a spherical shell defined by the CMB; that is, we evolve from an early-time basis into another different basis which is more convenient at late times.   We consider two late-time bispectra $A_{\ell_1\ell_2\ell_3}$ and $B_{\ell_1\ell_2\ell_3}$ depending on the spherical harmonic multipoles $\ell _i$ and we take the following inner product betweeen them:
\begin{align}
\<A,\, B\>_l &\equiv \sum_{\ell_i} \ \left( \frac{h_{\ell_1 \ell_2 \ell_3}}{v_{\ell_1}v_{\ell_2}v_{\ell_3}}\right)^2 \,A_{\ell_1 \ell_2 \ell_3} \, B_{\ell_1 \ell_2 \ell_3}\,,
\end{align}
where the required weight function is:
\begin{align}\label{eq:weight-function}
h^2_{\ell_1 \ell_2 \ell_3} = \frac{(2\ell_1+1)(2\ell_2+1)(2\ell_3+1)}{4\pi} \(\begin{array}{ccc}\ell_1 & \ell_2 & \ell_3 \\ 0 & 0 & 0 \end{array}  \)^2\,.
\end{align}
The array in brackets is the Wigner 3$J$ symbol, which is a geometric factor related to the projection onto the 2-sphere of the CMB.  It has two important properties. Firstly, it is zero if the three $\ell_i$, when treated as lengths, are unable to form a triangle (the triangle condition) and secondly it is zero whenever $\ell_1+\ell_2+\ell_3$ is odd (the parity condition). These two conditions present complications in evaluating the sums over the $l_i$ as the region is non-rectangular and inside the allowed region all odd combinations must be excluded. These two constraints present issues with load balancing and vectorisation, which will be discussed later.  When both these conditions are met, $h^2$ has an exact expression in terms of factorials which in turn can be accurately calculated using the Gosper approximation to the Wigner 3J symbol. The resulting expression is:
\begin{align}
h^2_{\ell_1 \ell_2 \ell_3} \approx \frac{1}{2\pi^2} \frac{(2\ell_1+1)(2\ell_2+1)(2\ell_3+1)(L+1/3)}{(L+1)(L_1+1/3)(L_2+1/3)(L_3+1/3)}\sqrt{\frac{(L_1+1/6)(L_2+1/6)(L_3+1/6)}{(L+1/6)}}
\end{align}
where $L=\ell_1+\ell_2+\ell_3$ and $L_i = L - 2\ell_i$.
This inner product needs to be calculated between our projected primordial basis $\widetilde{Q}$, which unavoidably involves a radial integral along a line of sight from primordial times until now, and our late time basis $Q$. For the estimation and projection to be tractable we must form our basis functions from products of one-dimensional basis functions $\tilde{q}$ and $q$ respectively which must be symmetrised over the $\ell_i$. Writing it out explicitly in terms of these 1D functions:
\begin{align}\label{eq:3D}
\nonumber \Gamma'_{nn'} = \<Q_n \widetilde{Q}_{n'} \> &= \sum_{\ell_1=2}^{\ell_{max}}\sum_{\ell_2=2}^{\ell_{max}}\sum_{\ell_3=2}^{\ell_{max}}  \frac{\h_{\ell_1\ell_2\ell_3}}{72\pi^2} \frac{(2\ell_1+1)(2\ell_2+1)(2\ell_3+1)(L+1/3)}{(L+1)(L_1+1/3)(L_2+1/3)(L_3+1/3)v_{\ell_1}v_{\ell_2}v_{\ell_3}} \\
\nonumber &\times \sqrt{\frac{(L_1+1/6)(L_2+1/6)(L_3+1/6)}{(L+1/6)C_{\ell_1}C_{\ell_2}C_{\ell_3}}} \(q_i(\ell_1)q_j(\ell_2)q_k(\ell_3) + 5\,\mbox{perms}\)\\
&\times  \int r^2dr \(\tilde{q}_{i'}(r,\ell_1) \tilde{q}_{j'}(r,\ell_2) \tilde{q}_{k'}(r,\ell_3) + 5\,\mbox{perms}\)
\end{align}
where $\h$ is a top-hat like function (which is 1 when the triangle and parity conditions are met and 0 elsewhere) and there is a known mapping between $n\rightarrow ijk$.  Evaluation of Equation~\ref{eq:3D} (where $v_\ell$, $C_\ell$ and $\tilde{q}$ have been precomputed) is performed by ``Modal3D'' (note that as we have a radial integral in addition to the three summations, the calculation is in fact four-dimensional for each of the $n_{max}^2$ independent matrix entries). For full details of the origin of Equation~\ref{eq:3D} see the Appendix. Typical values for the problem size are $n_{max}\px2000$, $\ell_{max}\px 2000$ and approximately $200$ points needed to calculate the integral over $r$ using splines.

\subsection{Separable Integration}
One significant simplification that can be made in some cases is to note that the Wigner 3$J$ symbol can be written as an integral over three Legendre polynomials so $h^2$ in Equation~\ref{eq:weight-function} takes the exact form:
\begin{align}
h^2_{\ell_1 \ell_2 \ell_3} = \frac{(2\ell_1+1)(2\ell_2+1)(2\ell_3+1)}{8\pi} \int d\mu P_{\ell_1}(\mu) P_{\ell_2}(\mu) P_{\ell_3}(\mu)\,.
\end{align}
This automatically preserves the triangle and parity conditions allowing us to work on a  domain which is both rectangular and dense.  This allows us to write Equation~\ref{eq:3D} in a much simpler form:
\begin{align}\label{eq:2D}
\Gamma'_{nn'} &= \frac{1}{48\pi} \int r^2 dr \int d\mu \(P_{ii'}(r,\mu)P_{jj'}(r,\mu)P_{kk'}(r,\mu) + 5\,perms\)\,,
\end{align}
where we have made the definition:
\begin{align}\label{eq:legendre}
P_{ii'}(r,\mu) &\equiv \sum_\ell \frac{(2\ell+1)}{v_\ell \sqrt{C_\ell}} \tilde{q}_{i'}(r,\ell) q_i(\ell) P_\ell(\mu)\,.
\end{align}
The simplified expression in Equation~\ref{eq:2D} is solved by ``Modal2D''.   One consequence of trying to retain the step like conditions of $h^2$ while using an integral form is that this calculation is very sensitive and must be done to very high precision.  Fortunately as the integrand is polynomial we can use Gauss-Legendre quadrature which is, in principle, exact.  However we found that standard libraries for calculation of weights and abscissas were not sufficiently accurate and we had to use a specialist implementation which had been optimised for this purpose \cite{hale2013fast} before the calculation proved stable for specific choices for $Q$ and $\bar{Q}$.  We note that this stability is not sufficient to allow us to reverse the order of the $r$ and $\mu$ integrations and the $\mu$ integration must always be carried out first.

It is this issue with stability which leads us to retaining both implementations of the calculation. The 2D version is employed for any bispectra where bases can be chosen so that the integral representation of $h^2$ is stable, with any remaining cases being calculated by the slower but robust 3D version.  We found that the majority of bispectra we are interested in can use the 2D version but there are some very well-motivated cases that cannot, for example the bispectrum induced by gravitational lensing.

The original code is written in C and parallelized with MPI. The 2D variant is parallelized over the loop of all the inner products $\<Q_n \widetilde{Q}_{n'} \>$, whereas the 3D variant is parallelized over the $n$ and $\ell_1$ indices in Equation \ref{eq:3D}.  Throughout the rest of this paper, ``Modal'' without a 2D or 3D suffix can be assumed to refer to both variants of the application. 

\subsection{\XeonPhi Coprocessors}

\noindent An \XeonPhi coprocessor consists of many ($\approx 60$) low frequency in-order cores which share a coherent memory.  Each of these cores can run up to four hardware threads, which is useful for hiding the latencies of memory accesses and multi-cycle instructions in the absence of out-of-order execution.  Unlike the ``hyperthreads'' on \Xeon processors, a single thread on the coprocesor cannot issue instructions on back-to-back cycles -- it is therefore necessary to use at least two threads per core to fully utilize the hardware.

Each core additionally has support for 512-bit SIMD instructions from the Initial Many-core Instruction (IMCI) set. Executing a fully vectorized fused multiply add (FMA) every cycle amounts to 16 double precision FLOPs per cycle per core, or a theoretical peak of $\approx$1 TFLOP/s in double precision.

The coprocessor is physically mounted on a PCIe card with its own GDDR memory and Linux operating system. Since the cores are x86-based and run Linux, the coprocessor is amenable to existing parallel programming languages and libraries such as MPI and OpenMP -- compiling existing codes for the coprocessor is often very simple, but extracting performance may require some significant algorithmic restructuring to expose sufficient parallelism~\cite{jeffers2013intel}.

\subsection{Experimental Setup}

\begin{table}
  \centering
  \caption{System configuration for a single node of \emph{Cosmic}.}
  \label{table:hardware}
  \setlength{\tabcolsep}{2pt} 
  \begin{tabular}{|r|c|c|}
    \hline
    & \textbf{\Xeon Processor E5-4650L} & \textbf{\XeonPhi Coprocessor 5110P} \\
    \hline	\hline
    {Sockets$\times$Cores$\times$Threads} & $1\times8\times1$ & $1\times60\times4$ \\
    {Clock (GHz)} & 2.6 & 1.053 \\
    {Double Precision GFLOP/s} & 166.4 & 1010.88 \\
    {L1 / L2 / L3 Cache (KB)} & 32 / 256 / 20,480 & 32 / 512 / - \\
    {DRAM (per node)} & 58 GB & 8 GB GDDR \\
    {STREAM~\cite{STREAM} Bandwidth} & 26.4 GB/s & 165 GB/s \\
    \hline
    \hline
    {PCIe Bandwidth} & \multicolumn{2}{c|}{6 GB/s} \\
    {Compiler Version} & \multicolumn{2}{c|}{icc v15.0.0.090} \\
    {MPI Version} & \multicolumn{2}{c|}{SGI MPT 2.10} \\
    {MPSS Version} & \multicolumn{2}{c|}{3.2.1} \\
    \hline
  \end{tabular}
\end{table}

\noindent We use the \emph{Cosmic} supercomputer at the University of Cambridge, which is an SGI UV2000 system consisting of 28 processors, 24 coprocessors, and 1.6 TB of RAM. The specification of a single node of \emph{Cosmic} is given in Table~\ref{table:hardware}.  The host processors are based on the microarchitecture previously code-named ``Sandy Bridge'', and the coprocessors on the microarchitecture previously code-named ``Knights Corner''.  For the sake of brevity, we refer to these two microarchitectures henceforth as ``SNB'' and ``KNC'', respectively.
 
All experiments (unless otherwise stated) were performed using all of the cores available within a node, running the maximum number of threads supported -- 4 threads on KNC and 1 thread on SNB (hyper-threading is disabled on the host) -- and the thread affinity used on KNC is set using: \texttt{KMP\_AFFINITY=\allowbreak close,granularity=fine}.  KNC is used in offload mode, and no additional flags are passed to the compiler beyond those used for SNB: \texttt{-O3 -xHost -mcmodel=medium -restrict -align -fno-alias -qopenmp}.  All experiments are repeated 5 times, and we present the average (mean), in order to account for system noise and any non-deterministic performance effects arising from threading.
 
Profiling and analysis of the code was performed using \VTuneTM and the optimisation reporting functionality of the \Intel C compiler.  For both forms of the algorithm, we measure the performance in terms of the number of loop iterations executed by the whole code per second.  Since the amount of work per loop iteration is different in the two cases, we differentiate between them by referring to them as 2D and 3D iterations per second (2D it/s and 3D it/s) respectively.  Henceforth we will also adopt a convenient notation for displaying the performance figures for SNB and KNC side-by-side as a tuple -- (SNB, KNC) it/s.

%% file: implementation.tex
\section{Optimisation and Modernisation}\label{sec:implementation}

\noindent Prior to exploring any algorithmic changes, hardware-specific optimisations or code modernisations, we carried out some high-level refactoring of the two Modal variants to improve their performance.  Specifically, we replaced a number of abstract ``getter'' functions used to retrieve array values from other compilation modules with inlined direct array accesses, removing function call overheads and providing the compiler with more optimisation freedom.  Some occurrences of routines from the GNU Scientific Library (GSL) were also replaced with equivalent, pre-optimised, routines from the \Intel Math Kernel Library (MKL).

The purpose of performing such a high-level refactoring before engaging in other optimisation activities is twofold: first, it ensures that the baseline performance of the code is representative of the performance of the original algorithm (as opposed to only its implementation); second, it exposes more accurately the incremental performance improvements from our subsequent changes to the code.  After our refactoring, the SNB performance of the 3D variant of the algorithm increased by 3.27\x, however the 2D variant showed negligible improvement.

\subsection{Modal2D}

\noindent The hotspot in the Modal2D code is in a function called \texttt{gamma\char`_pt} (see Listing~\ref{lst:gamma_pt_2D}) which evaluates the inner-most integral over $\mu$ in Equation~\ref{eq:2D-1} via Gauss-Legendre (GL) quadrature, which we repeat here for clarity;
\begin{align}\label{eq:2D-1}
\Gamma'_{nn'} &= \frac{1}{48\pi} \int r^2 dr \int d\mu \(P_{ii'}(r,\mu)P_{jj'}(r,\mu)P_{kk'}(r,\mu) + 5\,perms\) \\
P_{ii'}(r,\mu) &\equiv \sum_\ell \frac{(2\ell+1)}{v_\ell \sqrt{C_\ell}} \tilde{q}_{i'}(r,\ell) q_i(\ell) P_\ell(\mu)\,.
\end{align}
The indexes $mn$ are the element of $\Gamma'$ being calculated with the index $i$ corresponding to the point in the $r$ integration. The $j$ loop is over the GL-quadrature points for $\mu$, $s$ cycles over $ijk$ and $r$ over $i'j'k'$.  The array \texttt{Nmap} is $P_{ii'}$, with \texttt{beta} corresponding to $\tilde{q}$ and \texttt{basis} to $q$. The array \texttt{factor} contains the product of  all functions solely of $\ell$ and the Legendre polynomial $P_\ell$.  The $s1$-$6$ are the six permutations of the product of the $P_{ii'}$ required by symmetry and finally \texttt{gl\char`_wgt} is the GL-quadrature weight.  Since the size of $\Gamma$ is $\mathcal{O}(1000^2)$, there is a significant amount of exploitable parallelism present.

\begin{lstlisting}[style=prettyc,basicstyle=\footnotesize,caption={Code snippet for \texttt{gamma\_pt} in Modal2D.},label={lst:gamma_pt_2D}]
double gamma_pt(int m, int n, int i) {
	...
	for(j=0;j<npt;j++) {
		for(l=0;l<lsize;l++) {
			factor[l] =  lweight[l]*gl_Pl[j][l];
		}
		for(r=0;r<3;r++) {
			for(s=0;s<3;s++) {
				sum1 = 0.0;
				for(l=0;l<lsize;l++) {
					sum1 += factor[l]*basis[pvec[r]][l]*beta[i][qvec[s]][l];
				}
				Nmap[r][s] = sum1;
			}
		}
		s1  = Nmap[0][0]*Nmap[1][1]*Nmap[2][2];
		s2  = Nmap[0][0]*Nmap[1][2]*Nmap[2][1];
		s3  = Nmap[0][1]*Nmap[1][0]*Nmap[2][2];
		s4  = Nmap[0][1]*Nmap[1][2]*Nmap[2][0];
		s5  = Nmap[0][2]*Nmap[1][0]*Nmap[2][1];
		s6  = Nmap[0][2]*Nmap[1][1]*Nmap[2][0];
		
		sum1 = s1+s2+s3+s4+s5+s6;
		result += gl_wgt[j]*sum1;
	}
	...
	return result;
}
\end{lstlisting}

\subsubsection{Threading}

\noindent The original code has the loop over $mn$ points as a nested loop ($n$ as the outer loop, $m$ as the inner loop), and distributes the outer loop across processors using MPI.  We opt to extend this by distributing both the inner loop and the MPI subdomain of the outer loop across cores within a node using OpenMP.  Since the amount of work performed for each iteration of the $m$ loop is the same, and the calculation of each $mn$ point is separable and independent of all others, this is achieved simply through the application of an \texttt{omp parallel for collapse(2)} pragma.

\subsubsection{Vectorisation}
\noindent The code in its original form does not express the algorithm in a way that is conducive to compiler auto-vectorisation, and the \Intel compiler is unable to vectorise the $rsl$ loop nest because of the indirection arising from the \texttt{pvec} and \texttt{qvec} arrays.  Since the $rs$ loop nest has a hard-coded trip count of 9 iterations, it is practical to simply unroll this loop nest and compute the 9 sums simultaneously, thereby replacing the \texttt{pvec} and \texttt{qvec} with scalars and removing the indirection.  Loop unrolling also gives the compiler more freedom to re-order the instructions, which is particularly important for an in-order core like that of KNC.  Following these code transformations the compiler is able to auto-vectorise the loop, resulting in speed-ups of 2.65\x on SNB and 7.0\x on KNC -- performance of (5.2, 11.32) 2D it/s.  Tuning the data layout to assist vectorisation (\textit{i.e.} ensuring data alignment) and hoisting some repeated calculations from the inner-most loop increases performance further, to (5.3, 12.0) 2D it/s.

\subsubsection{Algorithmic Improvements}

\noindent A significant issue with the original algorithm is that the value $P^{T}_{ii'}(x,\mu)$  is calculated repeatedly from Equation~\ref{eq:legendre}.  Looking at the sizes of the dimensions of Equation~\ref{eq:legendre} -- $0 \le i \le \sqrt[3]{N_{terms}}$, $0 \le x \le 216$, and $0 \le \mu \le 3000$ -- if we were to pre-calculate $P_{ii'}$ for all values of $i$, $\mu$, and $x$ then the resultant array would require only 1~GB of storage (in double precision), which is small enough to remain within the 8~GB of GDDR available on KNC.  Pre-computing $P_{ii'}$ in this way reduces algorithmic complexity significantly, yielding a dramatic speed-ups of 278\x on KNC and 300\x on SNB -- giving a new performance of (1649.3, 3344.5) 2D it/s.

Following the introduction of the new algorithm, it is necessary to re-examine the hotspots and any assumptions about achievable performance.  Since the size of the look-up table we introduce is larger than the size of cache, and its access pattern is too irregular to benefit from blocking, our changes shift Modal2D from compute- to memory-bound.  Furthermore, running the full number of threads per core causes contention for entries in each core's Translation Look-aside Buffer (TLB) -- somewhat counter-intuitively, we can increase performance by a further 5\% by \emph{reducing} the number of threads used per core to 2.

\subsubsection{Results}

\begin{figure}
 \centering
 \includegraphics[width=0.7\linewidth]{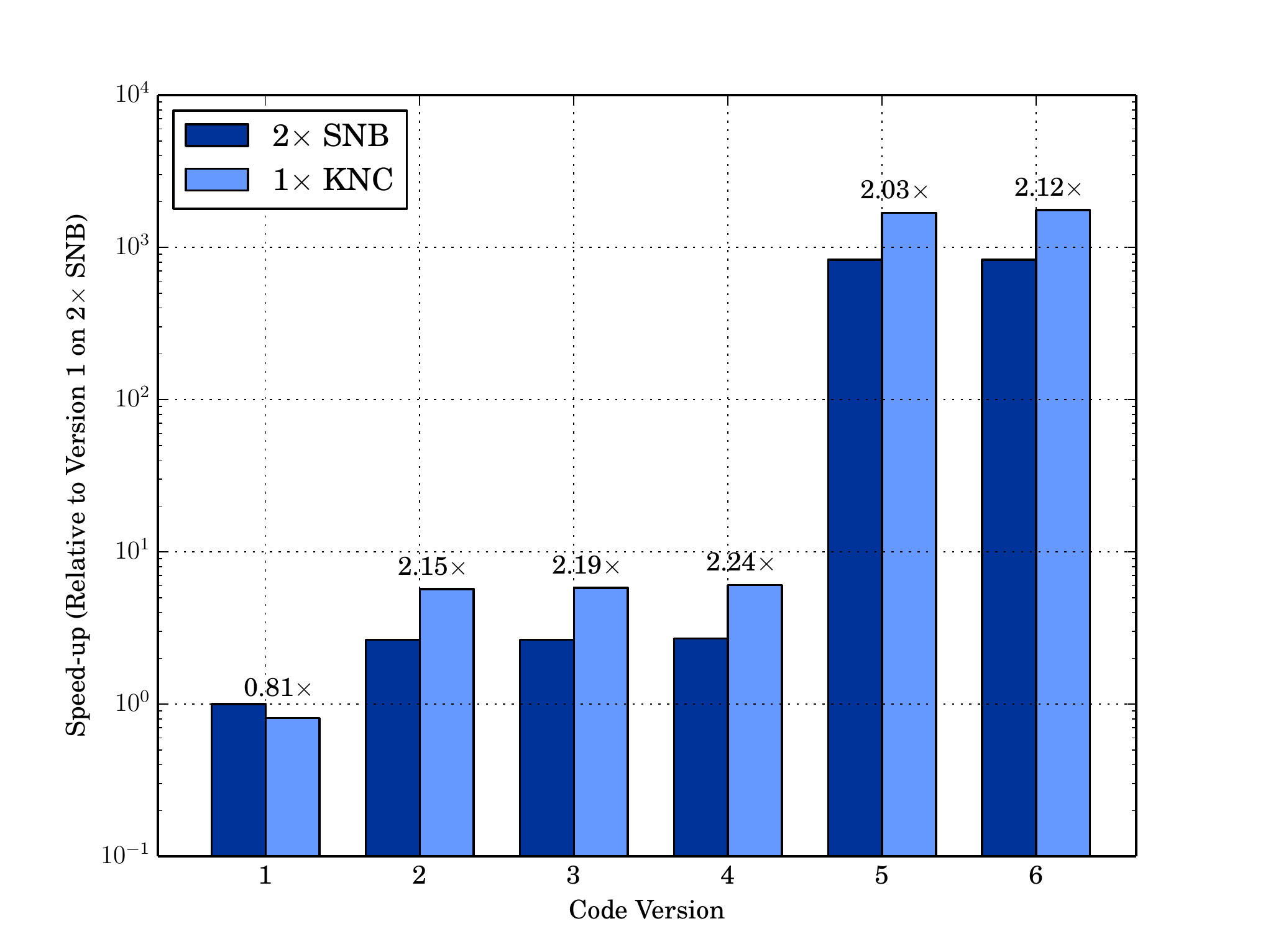}
 \caption{Performance improvement of Modal2D from optimisation and modernisation. The numbers above each of the bars are the performance difference between a coprocessor and two processors for that version.}
 \label{fig:performance-modal2d}
\end{figure}

\begin{table}
    \centering
    \caption{Execution times of Modal2D at various stages of code optimisation.}
    \label{table:2d-exe-time}
    \begin{tabular}{|l|l|l|l|}
	\hline
     \textbf{Version} & \textbf{2\x SNB (s)} & \textbf{1\x KNC (s)} & \textbf{Comment} \\
	\hline
        1     & $182005.17$  & $225028.22$  & Original code. \\
        2     & $68748.59$   & $31906.09$   & Unroll and vectorize loop. \\ 
        3     & $68628.19$   & $31304.09$   & Memory alignment. \\ 
        4     & $67303.79$   & $30100.08$   & Remove repeated calculations. \\ 
        5     & $219.0$      & $108.0$      & New algorithm. \\ 
        6     & $219.0$      & $103.33$     & Reduce number of threads per core to 2 on KNC. \\
	\hline
    \end{tabular}
\end{table}

\noindent The graph in \figurename~\ref{fig:performance-modal2d} shows the speed-up resulting from each of our optimisations and modernisations. For the original code following re-factoring and the introduction of threads (Version 1) we see that KNC is out-performed by two SNB sockets.  However, tuning to ensure efficient use of vectors (Versions 2--4) is sufficient to invert this relationship, and highlights the importance of achieving high SIMD efficiency on KNC.  The most significant speed-ups arise from algorithmic change (Version 5) on both processor and coprocessor. On KNC we also show that using less threads per core can give a boost in performance for this case (Version 6). (See Table \ref{table:2d-exe-time}.)

\subsection{Modal3D}

\noindent In the original Modal3D code we perform a brute force computation of\ Equation~\ref{eq:3D}, which we repeat here for clarity
\begin{align}
\nonumber \Gamma'_{nn'} = \<Q_n \widetilde{Q}_{n'} \> &= \sum_{\ell_1=2}^{\ell_{max}}\sum_{\ell_2=2}^{\ell_{max}}\sum_{\ell_3=2}^{\ell_{max}}  \frac{\h_{\ell_1\ell_2\ell_3}}{72\pi^2} \frac{(2\ell_1+1)(2\ell_2+1)(2\ell_3+1)(L+1/3)}{(L+1)(L_1+1/3)(L_2+1/3)(L_3+1/3)v_{\ell_1}v_{\ell_2}v_{\ell_3}} \\
\nonumber &\times \sqrt{\frac{(L_1+1/6)(L_2+1/6)(L_3+1/6)}{(L+1/6)C_{\ell_1}C_{\ell_2}C_{\ell_3}}} \(q_i(\ell_1)q_j(\ell_2)q_k(\ell_3) + 5\,\mbox{perms}\)\\
&\times  \int r^2dr \(\tilde{q}_{i'}(r,\ell_1) \tilde{q}_{j'}(r,\ell_2) \tilde{q}_{k'}(r,\ell_3) + 5\,\mbox{perms}\)
\end{align}
The majority of the time is spent in the loop structure shown in Listing~\ref{lst:gamma_3D}.  The outer loop(\texttt{n}) iterates over rows of $\Gamma$ as the computation of the projected primordial modes, $\tilde{Q}$, done by $\ell$-triple by \texttt{calculate\char`_xint} (which includes the $r$ integration) is the most expensive operation. All pre-factors in $\ell_i$, $L_i$ and $L$ are calculated and stored in \texttt{z} before we compute the inner product with all the late time modes (calculated by \texttt{plijk}) for the $\ell$-triple.  The constraints on the allowed $\ell$-triples give rise to the triangular \texttt{l1,l2,l3} loops (note the parity constraint in the \texttt{l3} loop which ensures that the sum of the $\ell_i$ is even).  

\begin{lstlisting}[style=prettyc,basicstyle=\footnotesize,caption={Code snippet for the hotspot in Modal3D.},label={lst:gamma_3D}]]
for(n=0; n<terms; n++) {
    for(l1=0; l1<lsize; l1++) {
        for(m=0; m<terms; m++) mvec[m] = 0.0;
        for(l2=l1; l2<lsize; l2++) {
            for(l3=l2+l1%2; l3<min(l1+l2,lmax)+1; l3+=2) {
                x = calculate_xint(l1,l2,l3,n,xsize,xvec,yvec,task);
                ...
                z = permsix(l1,l2,l3)*calculate_geometric(l1,l2,l3)/sqrt(s1*s2*s3);
                for(m=0;m<terms;m++) {
                    y = plijk(m,l1,l2,l3);
                    mvec[m] += x*y*z;
                }   
            }   
        }   
    }
    // array reduction of mvec into gamma matrix
}
\end{lstlisting}

\subsubsection{Threading}
 
\noindent As with Modal2D, the original code is parallelised with MPI only over the \texttt{n} loop, and introducing threading via OpenMP is a necessary step to improving utilisation of KNC's many-core architecture.  However, in the 3D variant there are multiple candidate loops that could be parallelised with OpenMP.  The inner-most \texttt{m} loop best matches the approach taken in the 2D case, but would be an unwise choice here (since the calculation of \texttt{x} and \texttt{z} would not benefit from parallelisation).  Threading the $\ell_1\ell_2\ell_3$ loops instead would be a better idea, if not for the fact that the loop space is triangular; threading only one of the loops would lead either to work imbalance or an iteration count too small to fully utilize all of the available cores.

Ideally we would like to flatten the whole $\ell_1\ell_2\ell_3$ space into a single loop, similar to the effect of OpenMP's \texttt{collapse(3)} construct.  Use of this construct is illegal here, since the number of iterations of the inner loops depends on the indices of the inner loops~\cite{openmp13}.  We experimented with many different approaches of achieving good load balance using standard OpenMP: spawning separate OpenMP tasks for groups of loop iterations; recursively partitioning the iteration space; and using dynamic scheduling with a threaded outer-loop.  In all cases we found that the overheads of these methodologies was significant, the load imbalance was not necessarily adequately addressed, and cache behaviour suffered due to the inability to specify thread affinities.  The \Intel Threading Building Blocks~\cite{reinders2007intel} (TBB) library provides task-affinity functionality that may have assisted us with this last point, but we have two reasons for not using it: first, to keep Modal as a C (rather than C++) application; and second, to ensure that our threaded version did not depend heavily on non-standard language features and libraries.

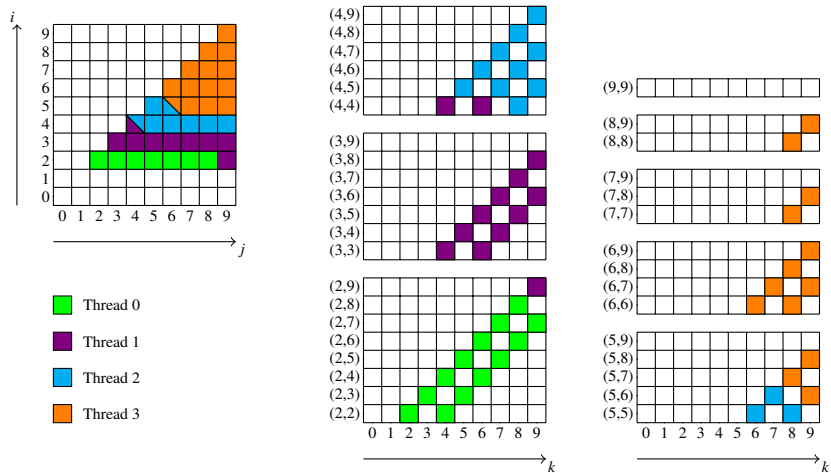
\begin{figure}
 \centering
 \input{plots/flattened.tex}
 \caption{Diagram explaining how we collapse the three loops into a single iteration space.}
 \label{fig:work-splitting}
\end{figure}

Instead, we carve up the iteration space manually across tasks, and assign contiguous ``chunks'' of loop iterations to each thread.  This scheme (demonstrated in \figurename~\ref{fig:work-splitting}) allows us to achieve very good load balancing at the expense of a small one-time setup overhead.  Note that this method is scalable beyond threads -- we can manually carve up the total $mn\ell_1\ell_2\ell_3$ iteration space across MPI, then across sockets/coprocessors and finally across threads -- which allows us to improve the multi-node scalability of the code.

While restructuring the loop for parallelism, we brought the $n$ loop inside of the $\ell_1\ell_2\ell_3$ nest to reduce the amount of redundant work per thread.  However, this increases the amount of temporary storage required by each thread to store its own partial result of $\Gamma_{mn}$. The size of this storage increases by the square of the number of terms, and is $\mathcal{O}(10\mbox{~MB})$ for a typical problem size.  Assuming 4 threads per core, the amount of temporary storage far exceeds the 512~KB capacity of KNC's L2 cache, and we see performance degradation due to thrashing effects.  If all the threads on one core \emph{collaborate} to read and write to the same array, then the size of this temporary storage is reduced by a factor of 4 -- while it still doesn't fit in the cache, blocking the loop to compute $B$ iterations between accesses can amortize the bandwidth overheads. 

Collaboration between threads in this way requires a ``nested'' parallelism model, with a 2-tier hierarchy of threads (\textit{e.g.} $C$ groups of 4 threads, where $C$ is the number of cores).  A na\"{i}ve implementation of nested parallelism in OpenMP is not practical here because the cost of creating and destroying new teams of threads for each iteration of the inner loop is prohibitively expensive.  Instead, we adopt an approach where the threads in both tiers are spawned together, and work is assigned to threads based on a combination of their team id (\textit{i.e} tid / 4) and local worker id (\textit{i.e.} tid \% 4) -- all threads execute the same set of $\ell_1\ell_2\ell_3$ loop iterations as the other threads in their group, but execute for different eigenmodes inside the angular momentum loop.

\begin{lstlisting}[style=prettyc,basicstyle=\footnotesize,caption={Example implementation of a fast, on-core, barrier.},label={lst:corebarrier}]]
typedef __declspec(align(64)) struct corebarrier_t
{
    volatile int barrier[2];
    int padding[14];
} corebarrier_t;

typedef __declspec(align(64)) struct barrier_t
{
    int barrier_id;    // which barrier to use this time (0 or 1)
    int core_tid;      // local thread id on the core (tid % 4)
    int flag[2];       // value to write this time (0 or 1)
    int waitval;       // value to wait on if threads are writing 1s
    corebarrier_t* me; // corebarrier_t shared by all threads on the core
} barrier_t;

void cpu_pause()
{
    // pause, sleep, or delay this thread
}

void corebarrier(barrier_t *bar)
{
    // determine which barrier and values to use
    int barrier_id = bar->barrier_id;
    int core_tid = bar->core_tid;
    int flag = bar->flag[barrier_id];
    int waitval = (flag) ? bar->waitval : 0;

    // loop until all threads have written to this barrier
    corebarrier_t *me = bar->me;
    ((char *)&me->barrier[barrier_id])[core_tid] = flag;
    while (me->barrier[barrier_id] != waitval)
        cpu_pause();

    // toggle the barrier/flag for next time
    bar->flag[barrier_id] = 1 - flag;
    bar->barrier_id = 1 - barrier_id;
}
\end{lstlisting}

Although this manual nesting approach gives us a great deal of flexibility, it depends on a custom on-core barrier (such as the one shown in Listing~\ref{lst:corebarrier}) to synchronize only the threads in a particular group.  There are two reasons to use a local barrier instead of the global barrier provided by OpenMP: first, a global barrier scales $O(log_2(4C))$, whereas a local barrier has a constant overhead of $O(4)$; and second, a global barrier can easily lead to accidental deadlocks, since all threads in all groups must be present for all barriers.

\subsubsection{Algorithmic Change}

\noindent The bottleneck left in the code after the loop modernisation is the integration routine called by \texttt{calculate\char`_xint}. This is a 1D integral of a function $f(r)$ with respect to the radial direction along the line of sight, $r$ from now back to the time of inflation. These functions are highly peaked at $r\approx 14,000$, around the time the CMB was formed, and are relatively flat everywhere else.  To decrease computational cost and memory consumption, these functions are sampled with 3 different resolutions in $r$, with more points around the surface of last recombination, and less points in the flat regions. Integration was performed by first fitting a spline to the data points and then integrating using the spline to interpolate points which are missing from the sample.  In the original code this procedure is performed by the routine \texttt{gsl\char`_spline\char`_eval\char`_integ} from the GSL, but was replaced with a faster equivalent from the \Intel MKL. We find that although the method from MKL has better vectorisation than the method from GSL, they both suffer from the same problems -- they are too memory intensive (storing a cubic spline in addition to the data) and computationally expensive (solving a set of linear equations).

We find that replacing this procedure with a simpler numerical integration routine can drastically speed-up the application even if it requires a greater number of sample points to achieve the same accuracy. We tried two integration methods for this - a simple application of the Trapezium rule and a method that uses Hermite Cubic spline interpolation. These two methods can compute the integral using local points only, unlike the GSL Spline which needs all points to compute the integral. Local methods are much more amenable to vectorisation and also requires $\mathcal{O}(1)$ temporary storage. The Hermite spline integrator is based on the interpolation of $r$ between points $r_k$ and $r_{k+1}$ using a Hermite cubic function:
\begin{equation}
 p(r_k) = (2t^3 - 3t^2 + 1)y_k + (t^3 - 2t^2 + t)(r_{k+1} - {r_k}) p'(r_k) + (-2t^3 + 3t^2) y_{k+1} + (t^3 - t^2) p'(r_{k+1}) \label{eq:hermite}
\end{equation}

\noindent Where $t = (r - r_k)/(r_{k+1} - r_k)$. The exact derivatives of $y(r)$ are not available so we must approximate. Here we use a simple approximation of $p'(r_k) = (y_{k+1} - y_{k}) / (x_{k+1} - x_{k})$. Integrating Equation~\ref{eq:hermite} with respect to $r$ and using this derivative yields the integrator:

\begin{equation}
 \int dr \; y(r) \approx \frac{1}{2}\sum_k \Delta r_k \left( y_k + y_{k+1} + \frac{1}{6}\Delta r_k \left(\frac{\Delta y_k}{\Delta r_k} - \frac{\Delta y_{k+1}}{\Delta r_{k+1}}\right)\right)
\end{equation}

For the results reported here, we use a simple application of the Trapezium rule, combined with 216 sampling points.  This scheme has sufficient numerical accuracy for us to obtain a physically meaningful answer, within a few percent of the answer calculated by GSL.  Other numerical integration routines (\textit{e.g.} Gaussian quadrature) may be required for other corner cases, and we leave this investigation to future work.

This leaves two bottlenecks in the code -- the partial reduction stage in \texttt{calculate\char`_gamma\char`_3d}, and the \texttt{calculate\char`_xint} routine. The reduction step is actually just a matrix multiply $\Gamma_{mn} = P_{lm}X_{ln}$, and thus it can be replaced with a call to the BLAS-3 DGEMM routine from \Intel MKL, which is cache-blocked and vectorized efficiently out-of-the-box.  We call MKL from only a single thread in each of our nested thread groups, and empirically derive the best blocking factor $B$ to be 64.

\subsubsection{Results}

\begin{figure}
 \centering
 \includegraphics[width=0.7\linewidth]{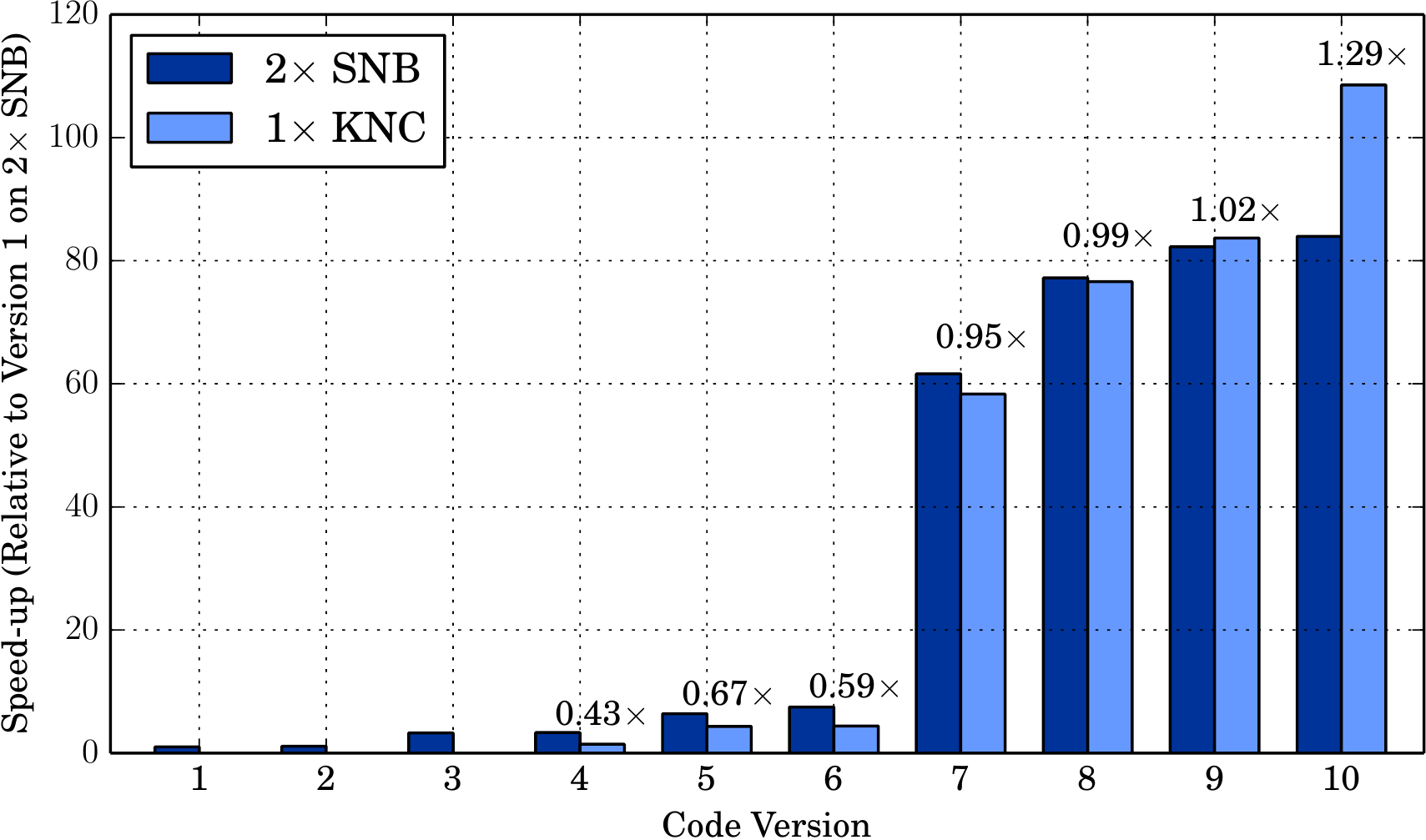}
 \caption{Performance improvement of Modal3D from optimisation and modernisation. The numbers above each of the bars are the performance difference between a coprocessor and two processors for that version.}
 \label{fig:performance-modal3d}
\end{figure}

\begin{table}
    \centering
    \caption{Execution times of Modal3D at various stages of code optimisation.}
    \label{table:3d-exe-time}
    \begin{tabular}{|l|l|l|l|}
	\hline
     \textbf{Version} & \textbf{2\x SNB (s)} & \textbf{1\x KNC (s)} & \textbf{Comment} \\
	\hline
        1     & 2887.0    &  -   & Original Code. \\
        2     & 2610.0    &  -   & Loop simplification.\\ 
        3     & 882.0     &  -   & MKL Integration. Remove getter methods.\\ 
        4     & 865.9 &  1991.6  & Flattened loops + threads.\\ 
        5     & 450.6   &  667.9 & Loop reorder, manual nested threading.\\ 
        6     & 385.6  &  655.0  & Blocked version of loop.\\
        7     & 46.9   &  49.5   & Trapezium Integration.\\
        8     & 37.4    &  37.7  & Reduction with DGEMM.\\
        9     & 35.1    &  34.5  & Alignment+Padding of Arrays.\\
        10    & 34.3    &  26.6  & Software Prefetching.\\
	\hline
    \end{tabular}
\end{table}

\noindent The graph in \figurename~\ref{fig:performance-modal3d} shows the speed-up resulting from each of our optimisations and modernisations.  For the original code following re-factoring and the introduction of threads (Version 1-3) we already see significant gains on the host (3.27\x).  Note that we do not provide KNC results until the introduction of OpenMP (Version 4) since we are using the offload model, and offloading work to a single KNC thread does not make sense.  Tuning the threading behaviour (Versions 5 and 6) continues to improve performance on both platforms but, as in the case of Modal2D, more significant speed-ups are only possible with algorithmic change -- in this case, a change in integration method (Version 7).  Further tuning of vectorisation behaviour (Versions 8 and 9) provide a small boost in performance, and tuning the prefetch distance (Version 10) finally pushes the performance of KNC ahead of SNB. (See Table \ref{table:3d-exe-time}.)

Since the same code can be recompiled and ran on both processor and coprocessor, very little effort was required to make the code heterogeneous.  We use the asynchronous offload features of \Intel Language Extensions for Offload (LEO) to offload a fraction (empirically derived to be 71\%) of the total problem to KNC, and process the remainder on SNB.  On a single node of \emph{Cosmic}, using KNC in addition to SNB gives a 3.5\x speed-up over using SNB only.

%% file: plots/flattened.tex
\begin{tikzpicture}[scale=0.24, transform shape, font=\Huge]

\begin{scope}
    \filldraw[fill=thread3,draw=black] (0,0) rectangle node[anchor=west,right=1.0cm] {Thread 3} ++ (1,1);
    \filldraw[fill=thread2,draw=black] (0,2) rectangle node[anchor=west,right=1.0cm] {Thread 2} ++ (1,1);
    \filldraw[fill=thread1,draw=black] (0,4) rectangle node[anchor=west,right=1.0cm] {Thread 1} ++ (1,1);
    \filldraw[fill=thread0,draw=black] (0,6) rectangle node[anchor=west,right=1.0cm] {Thread 0} ++ (1,1);
\end{scope}

\begin{scope}[yshift=12cm]
    \draw[draw=black,step=1cm] (0,0) grid (10,10);

    \filldraw[fill=thread0,draw=black] (2,2) rectangle (3,3);
    \filldraw[fill=thread0,draw=black] (3,2) rectangle (4,3);
    \filldraw[fill=thread0,draw=black] (4,2) rectangle (5,3);
    \filldraw[fill=thread0,draw=black] (5,2) rectangle (6,3);
    \filldraw[fill=thread0,draw=black] (6,2) rectangle (7,3);
    \filldraw[fill=thread0,draw=black] (7,2) rectangle (8,3);
    \filldraw[fill=thread0,draw=black] (8,2) rectangle (9,3);
    \filldraw[fill=thread1,draw=black] (9,2) rectangle (10,3);

    \filldraw[fill=thread1,draw=black] (3,3) rectangle (4,4);
    \filldraw[fill=thread1,draw=black] (4,3) rectangle (5,4);
    \filldraw[fill=thread1,draw=black] (5,3) rectangle (6,4);
    \filldraw[fill=thread1,draw=black] (6,3) rectangle (7,4);
    \filldraw[fill=thread1,draw=black] (7,3) rectangle (8,4);
    \filldraw[fill=thread1,draw=black] (8,3) rectangle (9,4);
    \filldraw[fill=thread1,draw=black] (9,3) rectangle (10,4);

    \filldraw[fill=thread1,draw=black] (4,4) -- ++(1,0) -- ++(-1,1) -- ++(0,-1);
    \filldraw[fill=thread2,draw=black] (4,5) -- ++(1,0) -- ++(0,-1) -- ++(-1,1);
    \filldraw[fill=thread2,draw=black] (5,4) rectangle (6,5);
    \filldraw[fill=thread2,draw=black] (6,4) rectangle (7,5);
    \filldraw[fill=thread2,draw=black] (7,4) rectangle (8,5);
    \filldraw[fill=thread2,draw=black] (8,4) rectangle (9,5);
    \filldraw[fill=thread2,draw=black] (9,4) rectangle (10,5);

    \filldraw[fill=thread2,draw=black] (5,5) rectangle (6,6);
    \filldraw[fill=thread2,draw=black] (6,5) -- ++(1,0) -- ++(-1,1) -- ++(0,-1);
    \filldraw[fill=thread3,draw=black] (6,6) -- ++(1,0) -- ++(0,-1) -- ++(-1,1);
    \filldraw[fill=thread3,draw=black] (7,5) rectangle (9,6);
    \filldraw[fill=thread3,draw=black] (8,5) rectangle (9,6);
    \filldraw[fill=thread3,draw=black] (9,5) rectangle (10,6);

    \filldraw[fill=thread3,draw=black] (6,6) rectangle (7,7);
    \filldraw[fill=thread3,draw=black] (7,6) rectangle (8,7);
    \filldraw[fill=thread3,draw=black] (8,6) rectangle (9,7);
    \filldraw[fill=thread3,draw=black] (9,6) rectangle (10,7);

    \filldraw[fill=thread3,draw=black] (7,7) rectangle (8,8);
    \filldraw[fill=thread3,draw=black] (8,7) rectangle (9,8);
    \filldraw[fill=thread3,draw=black] (9,7) rectangle (10,8);

    \filldraw[fill=thread3,draw=black] (8,8) rectangle (9,9);
    \filldraw[fill=thread3,draw=black] (9,8) rectangle (10,9);

    \filldraw[fill=thread3,draw=black] (9,9) rectangle (10,10);

    \foreach \x in {0,1,2,3,4,5,6,7,8,9}
        \draw[xshift=0.5cm] (\x cm,1pt) -- (\x cm,-1pt) node[anchor=north,below=0.1cm] {$\x$};
    \foreach \y in {0,1,2,3,4,5,6,7,8,9}
        \draw[yshift=0.5cm] (-1pt,\y cm) -- (1pt,\y cm) node[anchor=east,left=0.1cm] {$\y$};

    \draw[->] (0,-2) -- (10,-2) node[anchor=north west] {$j$};
    \draw[->] (-2,0) -- (-2,10) node[anchor=south east] {$i$};
\end{scope}

\begin{scope}[xshift=17cm]
    \foreach \x in {0,1,2,3,4,5,6,7,8,9}
        \draw[xshift=0.5cm] (\x cm,1pt) -- (\x cm,-1pt) node[anchor=north,below=0.1cm] {$\x$};
    \draw[->] (0,-2) -- (10,-2) node[anchor=north west] {$k$};
\end{scope}
\begin{scope}[xshift=32cm]
    \foreach \x in {0,1,2,3,4,5,6,7,8,9}
        \draw[xshift=0.5cm] (\x cm,1pt) -- (\x cm,-1pt) node[anchor=north,below=0.1cm] {$\x$};
    \draw[->] (0,-2) -- (10,-2) node[anchor=north west] {$k$};
\end{scope}


\begin{scope}[xshift=17cm]
    \draw[draw=black,step=1cm] (0,0) grid (10,8); 
    \filldraw[fill=thread0,draw=black] (2,0) rectangle (3,1);
    \filldraw[fill=thread0,draw=black] (4,0) rectangle (5,1);
    \filldraw[fill=thread0,draw=black] (3,1) rectangle (4,2);
    \filldraw[fill=thread0,draw=black] (5,1) rectangle (6,2);
    \filldraw[fill=thread0,draw=black] (4,2) rectangle (5,3);
    \filldraw[fill=thread0,draw=black] (6,2) rectangle (7,3);
    \filldraw[fill=thread0,draw=black] (5,3) rectangle (6,4);
    \filldraw[fill=thread0,draw=black] (7,3) rectangle (8,4);
    \filldraw[fill=thread0,draw=black] (6,4) rectangle (7,5);
    \filldraw[fill=thread0,draw=black] (8,4) rectangle (9,5);
    \filldraw[fill=thread0,draw=black] (7,5) rectangle (8,6);
    \filldraw[fill=thread0,draw=black] (9,5) rectangle (10,6);
    \filldraw[fill=thread0,draw=black] (8,6) rectangle (9,7);
    \filldraw[fill=thread1,draw=black] (9,7) rectangle (10,8);
    \foreach \y in {2,3,4,5,6,7,8,9}
        \draw[yshift=-1.5cm] (-1pt,\y cm) -- (1pt,\y cm) node[anchor=east,left=0.1cm] {(2,$\y$)};
\end{scope}

\begin{scope}[xshift=17cm,yshift=9cm]
    \draw[draw=black,step=1cm] (0,0) grid (10,7); 
    \filldraw[fill=thread1,draw=black] (4,0) rectangle (5,1);
    \filldraw[fill=thread1,draw=black] (6,0) rectangle (7,1);
    \filldraw[fill=thread1,draw=black] (5,1) rectangle (6,2);
    \filldraw[fill=thread1,draw=black] (7,1) rectangle (8,2);
    \filldraw[fill=thread1,draw=black] (6,2) rectangle (7,3);
    \filldraw[fill=thread1,draw=black] (8,2) rectangle (9,3);
    \filldraw[fill=thread1,draw=black] (7,3) rectangle (8,4);
    \filldraw[fill=thread1,draw=black] (9,3) rectangle (10,4);
    \filldraw[fill=thread1,draw=black] (8,4) rectangle (9,5);
    \filldraw[fill=thread1,draw=black] (9,5) rectangle (10,6);
    \foreach \y in {3,4,5,6,7,8,9}
        \draw[yshift=-2.5cm] (-1pt,\y cm) -- (1pt,\y cm) node[anchor=east,left=0.1cm] {(3,$\y$)};
\end{scope}

\begin{scope}[xshift=17cm,yshift=17cm]
    \draw[draw=black,step=1cm] (0,0) grid (10, 6); 
    \filldraw[fill=thread1,draw=black] (4,0) rectangle ++ (1,1);
    \filldraw[fill=thread1,draw=black] (6,0) rectangle ++ (1,1);
    \filldraw[fill=thread2,draw=black] (8,0) rectangle ++ (1,1);
    \filldraw[fill=thread2,draw=black] (5,1) rectangle ++ (1,1);
    \filldraw[fill=thread2,draw=black] (7,1) rectangle ++ (1,1);
    \filldraw[fill=thread2,draw=black] (9,1) rectangle ++ (1,1);
    \filldraw[fill=thread2,draw=black] (6,2) rectangle ++ (1,1);
    \filldraw[fill=thread2,draw=black] (8,2) rectangle ++ (1,1);
    \filldraw[fill=thread2,draw=black] (7,3) rectangle ++ (1,1);
    \filldraw[fill=thread2,draw=black] (9,3) rectangle ++ (1,1);
    \filldraw[fill=thread2,draw=black] (8,4) rectangle ++ (1,1);
    \filldraw[fill=thread2,draw=black] (9,5) rectangle ++ (1,1);
    \foreach \y in {4,5,6,7,8,9}
        \draw[yshift=-3.5cm] (-1pt,\y cm) -- (1pt,\y cm) node[anchor=east,left=0.1cm] {(4,$\y$)};
\end{scope}

\begin{scope}[xshift=32cm]
    \draw[draw=black,step=1cm] (0,0) grid (10, 5); 
    \filldraw[fill=thread2,draw=black] (6,0) rectangle ++ (1,1);
    \filldraw[fill=thread2,draw=black] (8,0) rectangle ++ (1,1);
    \filldraw[fill=thread2,draw=black] (7,1) rectangle ++ (1,1);
    \filldraw[fill=thread3,draw=black] (9,1) rectangle ++ (1,1);
    \filldraw[fill=thread3,draw=black] (8,2) rectangle ++ (1,1);
    \filldraw[fill=thread3,draw=black] (9,3) rectangle ++ (1,1);
    \foreach \y in {5,6,7,8,9}
        \draw[yshift=-4.5cm] (-1pt,\y cm) -- (1pt,\y cm) node[anchor=east,left=0.1cm] {(5,$\y$)};
\end{scope}

\begin{scope}[xshift=32cm,yshift=6cm]
    \draw[draw=black,step=1cm] (0,0) grid (10, 4); 
    \filldraw[fill=thread3,draw=black] (6,0) rectangle ++ (1,1);
    \filldraw[fill=thread3,draw=black] (8,0) rectangle ++ (1,1);
    \filldraw[fill=thread3,draw=black] (7,1) rectangle ++ (1,1);
    \filldraw[fill=thread3,draw=black] (9,1) rectangle ++ (1,1);
    \filldraw[fill=thread3,draw=black] (8,2) rectangle ++ (1,1);
    \filldraw[fill=thread3,draw=black] (9,3) rectangle ++ (1,1);
    \foreach \y in {6,7,8,9}
        \draw[yshift=-5.5cm] (-1pt,\y cm) -- (1pt,\y cm) node[anchor=east,left=0.1cm] {(6,$\y$)};
\end{scope}

\begin{scope}[xshift=32cm,yshift=11cm]
    \draw[draw=black,step=1cm] (0,0) grid (10, 3); 
    \filldraw[fill=thread3,draw=black] (8,0) rectangle ++ (1,1);
    \filldraw[fill=thread3,draw=black] (9,1) rectangle ++ (1,1);
    \foreach \y in {7,8,9}
        \draw[yshift=-6.5cm] (-1pt,\y cm) -- (1pt,\y cm) node[anchor=east,left=0.1cm] {(7,$\y$)};
\end{scope}

\begin{scope}[xshift=32cm,yshift=15cm]
    \draw[draw=black,step=1cm] (0,0) grid (10, 2); 
    \filldraw[fill=thread3,draw=black] (8,0) rectangle ++ (1,1);
    \filldraw[fill=thread3,draw=black] (9,1) rectangle ++ (1,1);
    \foreach \y in {8,9}
        \draw[yshift=-7.5cm] (-1pt,\y cm) -- (1pt,\y cm) node[anchor=east,left=0.1cm] {(8,$\y$)};
\end{scope}

\begin{scope}[xshift=32cm,yshift=18cm]
    \draw[draw=black,step=1cm] (0,0) grid (10, 1); 
    \foreach \y in {9}
        \draw[yshift=-8.5cm] (-1pt,\y cm) -- (1pt,\y cm) node[anchor=east,left=0.1cm] {(9,$\y$)};
\end{scope}

\end{tikzpicture}

%% file: results.tex
\section{Performance Study}\label{sec:results}

\noindent The new implementations of the 2D and 3D variants of Modal are \emph{optimised} but they are not \emph{optimal}.  In this section, we analyze and discuss the remaining bottlenecks to performance at both single- and multi-node scale.

\subsection{Integrator Performance: Accuracy vs. Execution Time}
In Modal3D, swapping out the spline-based integrator for a simpler one significantly accelerates the code on modern architectures, but at the price of reduce accuracy. We show that this accuracy can be both recovered and increased when using the simpler integrators by increasing the number of sampling points. Moreover, even with a larger sampling size the performance is still likely to be far better than the spline based approach using less points.

\begin{figure}

 \centering
 
 \begin{subfigure}[t]{0.48\linewidth}
  \includegraphics[width=\textwidth]{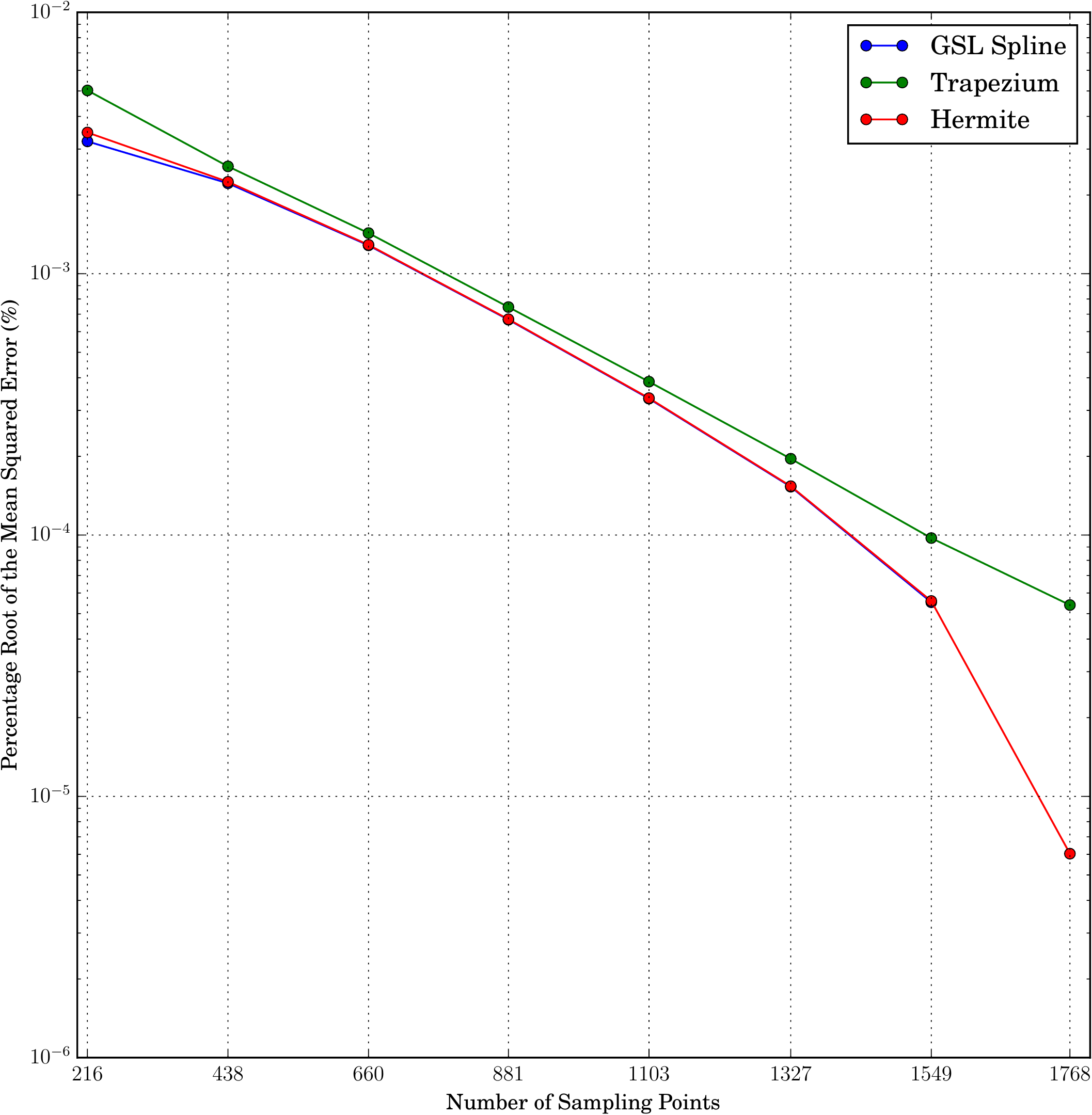}
  \caption{Percentage RMSE of $\Gamma$ using different integrators and numbers of sampling points compared to the GSL Spline method with 1768 sampling points.}
  \label{fig:integrator-error}
 \end{subfigure} %
 \hfill %
 \begin{subfigure}[t]{0.48\linewidth}
  \includegraphics[width=\textwidth]{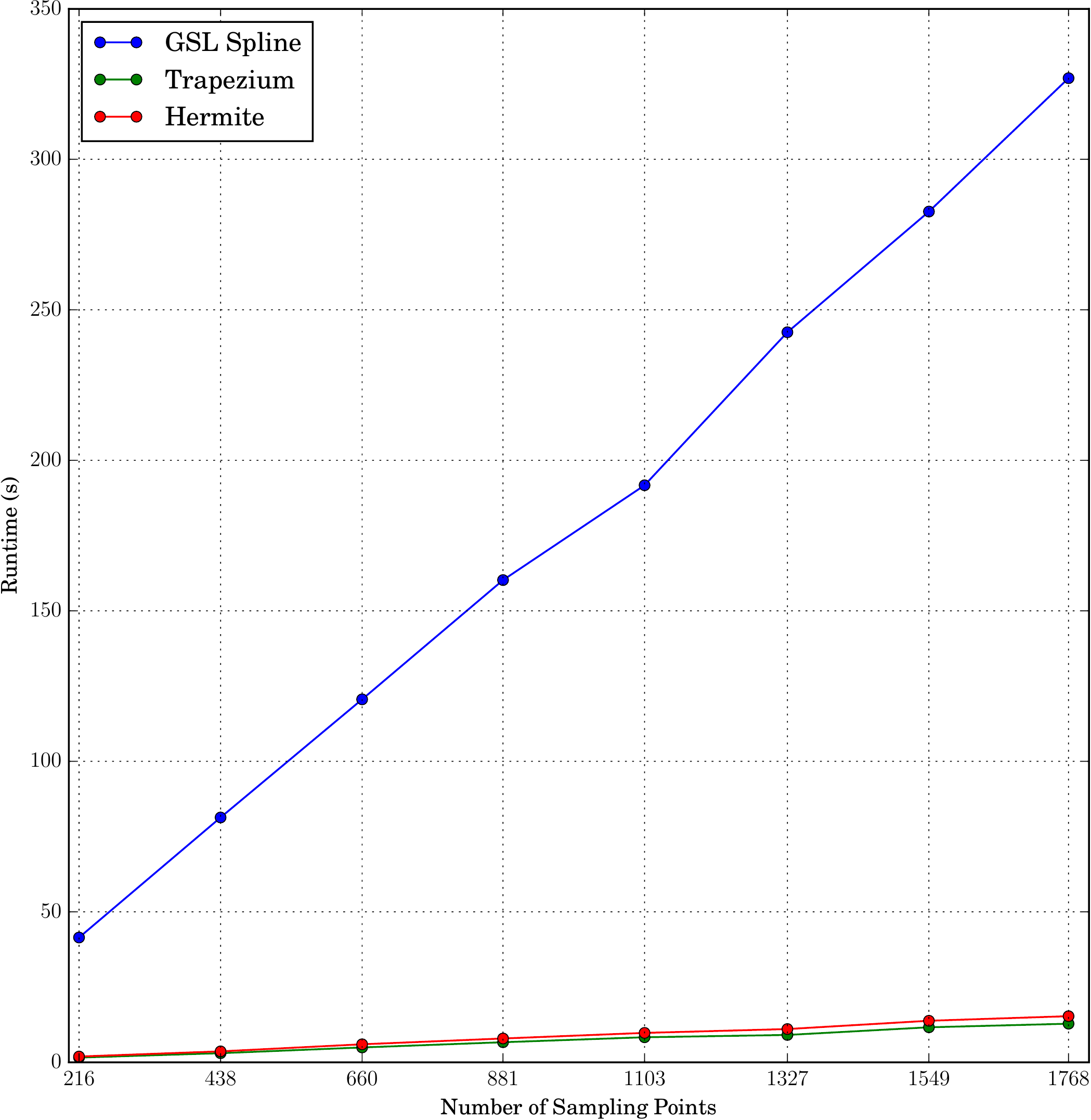}
  \caption{Runtimes of entire Modal3D calculation using different integrators and varying the numbers of sampling points.}
  \label{fig:integrator-times}
 \end{subfigure}
 
 \caption{Comparison of the accuracy and performance of the different integration methods used in Modal3D.}
 \label{fig:integrator-comparison}
 
\end{figure}

Figure~\ref{fig:integrator-error} shows the percentage root mean squared error (RMSE) of the unit normalised $\Gamma$ produced by Modal3D, for different integration methods and different numbers of sample points. We take the result of Modal3D with using the GSL Spline with 1768 sampling points to be the ``gold standard'' to compare all the others to. All three methods converge towards the gold standard with increasing sample points. The Hermite integrator has nearly identical error to the GSL spline and achieves an error of $<1\times10^{-5}\%$ of the gold standard with the maximum number of points. The Trapezium integrator is the least accurate of the three, but still gets an error of $<1\times10^{-4}\%$ of the gold standard. If we consider the accuracies in light of the times to solution shown in Figure~\ref{fig:integrator-times}, a clear picture emerges -- even after increasing the accuracy of the Hermite and Trapezium integrators, they remain much faster than the GSL spline with a small number of points. The Trapezium integrator is the fastest of the three, but is only $\sim1.25\times$ faster than the Hermite.

\subsection{Scaling with Cores}

\begin{figure}

 \centering
 
 \begin{subfigure}[b]{0.48\linewidth}
  \includegraphics[width=\textwidth]{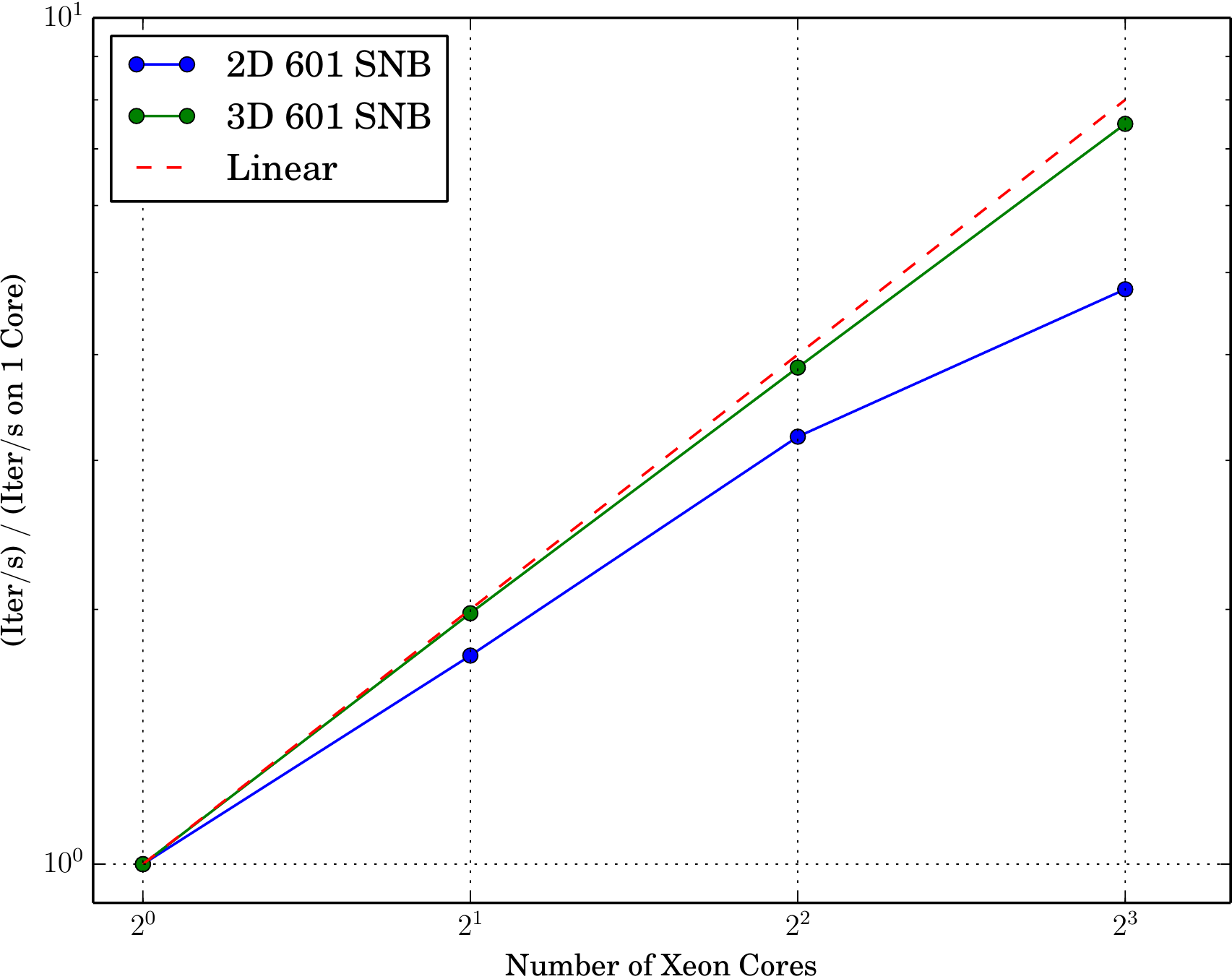}
  \caption{Modal2D}
  \label{fig:core-scaling-snb}
 \end{subfigure} %
 \hfill %
 \begin{subfigure}[b]{0.48\linewidth}
  \includegraphics[width=\textwidth]{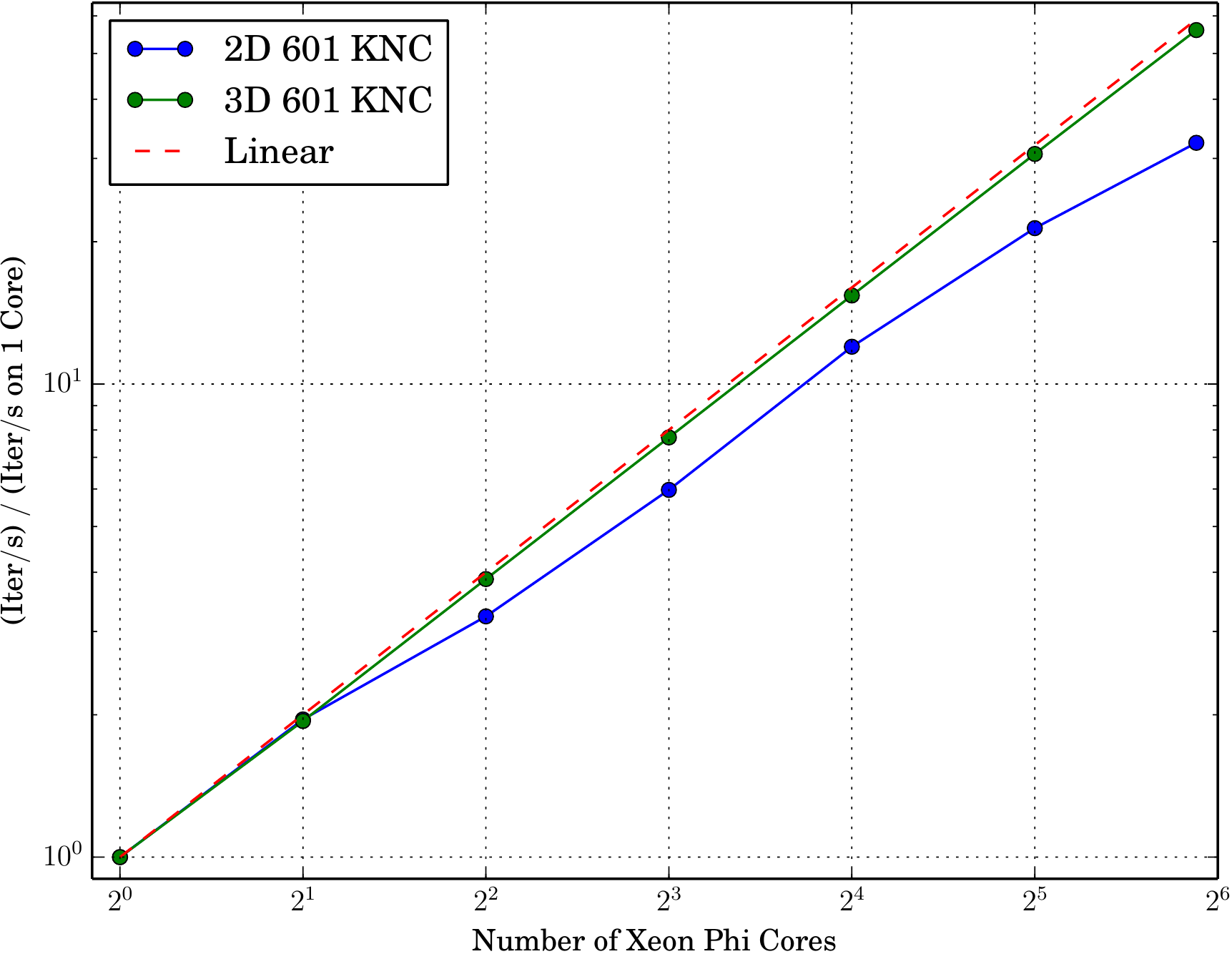}
  \caption{Modal3D}
  \label{fig:core-scaling-knc}
 \end{subfigure}
 
 \caption{Strong-scaling within a single node, for 601 modes.}
 \label{fig:core-scaling}
 
\end{figure}

\noindent Figures \ref{fig:core-scaling-snb} and \ref{fig:core-scaling-knc} show how performance of Modal2D and Modal3D scale with the number of cores on SNB and KNC, respectively.  The number of threads per core on KNC is fixed at the number that gives the highest performance in each case -- 2 and 4 threads per core for Modal2D and Modal3D, respectively.

For Modal2D, we see that scaling tapers off with increasing core-count, reaching a maximum speed-up of 32.4\x.  This is not for want of parallel work, nor high synchronisation costs, but rather because the cores are competing for limited memory bandwidth.  Running with all 59 cores on KNC, the bandwidth from GDDR to L2 was measured (by \VTune) to be 147.5 GB/s on average -- very close to its peak STREAM~\cite{STREAM} bandwidth of 165 GB/s (see Table~\ref{table:hardware}).  The fact that we are hitting close to peak shows that there is little room left to further tune the computation performed by Modal2D; any additional work on the 2D variant will need to focus on improving cache locality or algorithmic complexity, not vectorisation.

For Modal3D, we see much better scaling with increasing core-count (close to linear).  This is to be expected, since the code is very compute intensive.  Furthermore the groups of threads, once spawned, require no communication with each other until the reduction stage at the very end of the calculation.  The scaling behaviour suggests that, unlike the 2D variant, Modal3D is not bound by memory bandwidth.  However, we are not yet instruction bound, either -- running with all 59 cores on KNC, the number of vector instructions issued per cycle was measured by Speedometer~\cite{Speedometer} to be 39.8\% of peak.  The performance of Modal3D is in fact limited by transfers between L2 and L1 cache; although there is a large amount of data re-use within a core, the number of streams per thread is too high for the prefetchers to effectively hide the latency of L2 accesses.

\subsection{Scaling with Nodes}

\begin{figure}

 \centering
 
 \begin{subfigure}[b]{0.48\linewidth}
  \includegraphics[width=\textwidth]{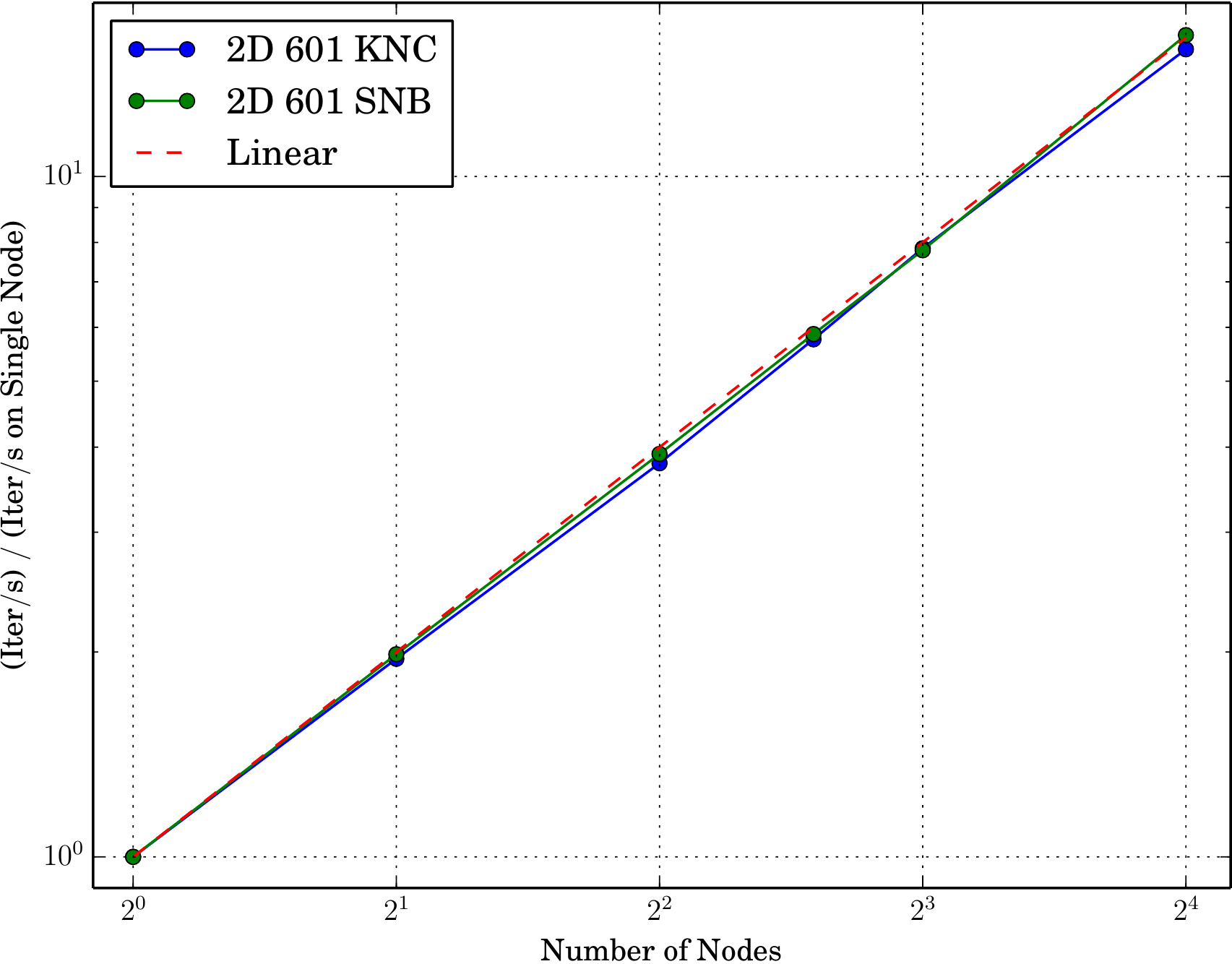}
  \caption{Modal2D}
  \label{fig:node-scaling-2d}
 \end{subfigure} %
 \hfill %
 \begin{subfigure}[b]{0.48\linewidth}
  \includegraphics[width=\textwidth]{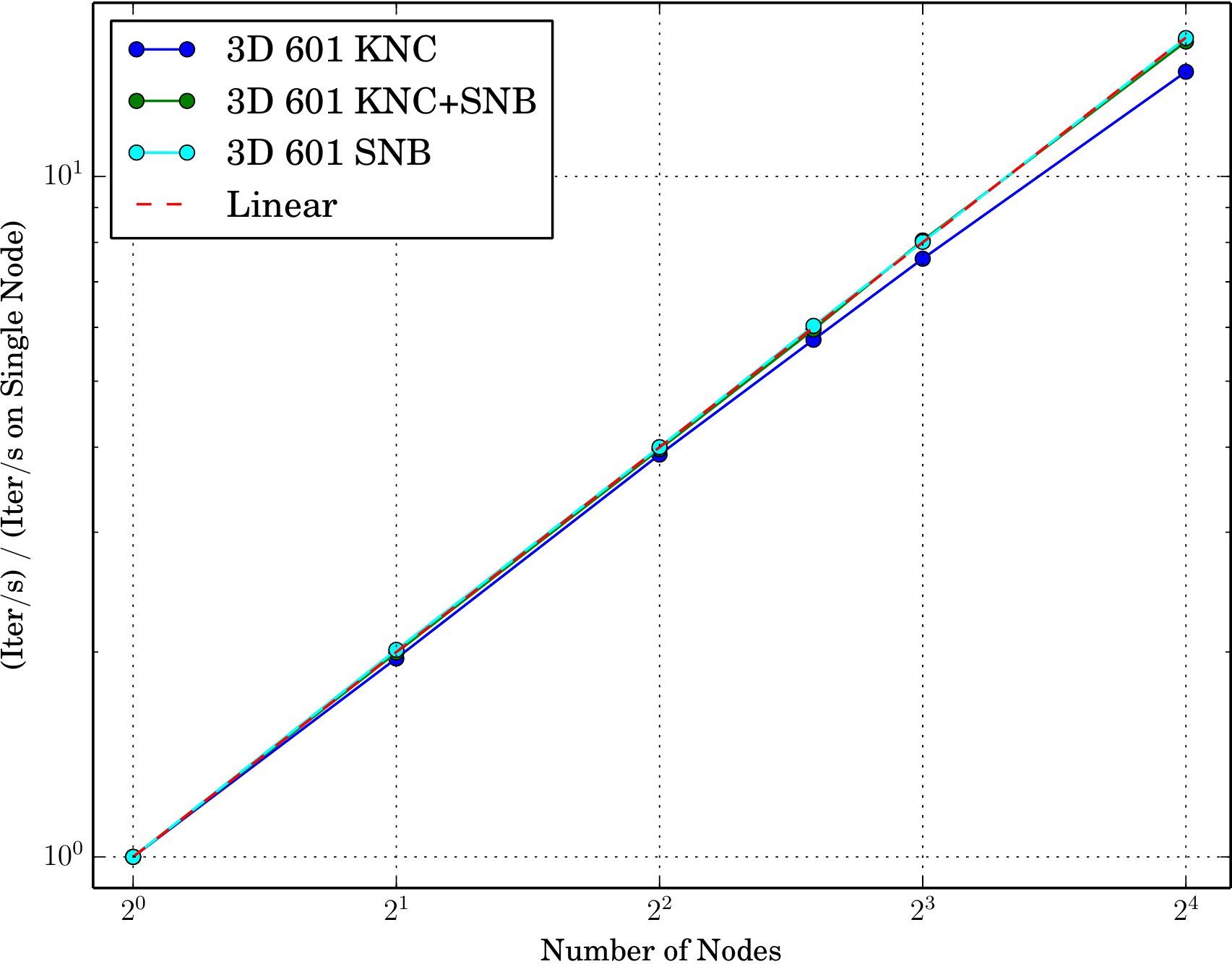}
  \caption{Modal3D}
  \label{fig:node-scaling-3d}
 \end{subfigure}
 
 \caption{Strong-scaling on multiple nodes, for 601 modes.}
 \label{fig:node-scaling}
 
\end{figure}

\noindent Figures \ref{fig:node-scaling-2d} and \ref{fig:node-scaling-3d} show how performance scales with the number of \emph{Cosmic} nodes for Modal2D and Modal3D, respectively.  For Modal2D we show results for SNB and KNC alone, while for Modal3D we also show results for a ``hybrid'' model running 30\% of the work on SNB and 70\% of the work on KNC.  In all cases, we place a single MPI rank per node.  The reader is reminded that a single \emph{Cosmic} node contains only a single SNB socket, and a single KNC coprocessor (see Table~\ref{table:hardware}).

For both the 2D and 3D variants, we see scaling that is fairly close to linear.  There are two reasons for this: first, the only communication required between tasks occurs right at the start of a run (to agree on task decomposition) and right at the end (to reduce the final gamma matrix); second, there is a significant amount of work (601 iterations in the 2D case, and $2\times 10^9$ iterations in the 3D case) to be split between MPI ranks, so we do not reach a scale where communication costs begin to dominate execution time.  Note that although the single-node scaling of Modal2D is inhibited by hardware limitations, this is not the case here -- when scaling to multiple nodes, the total available memory bandwidth also scales accordingly.

\subsection{Scaling with Problem Size}

\begin{figure}

 \centering
 
 \begin{subfigure}[b]{0.48\linewidth}
  \includegraphics[width=\textwidth]{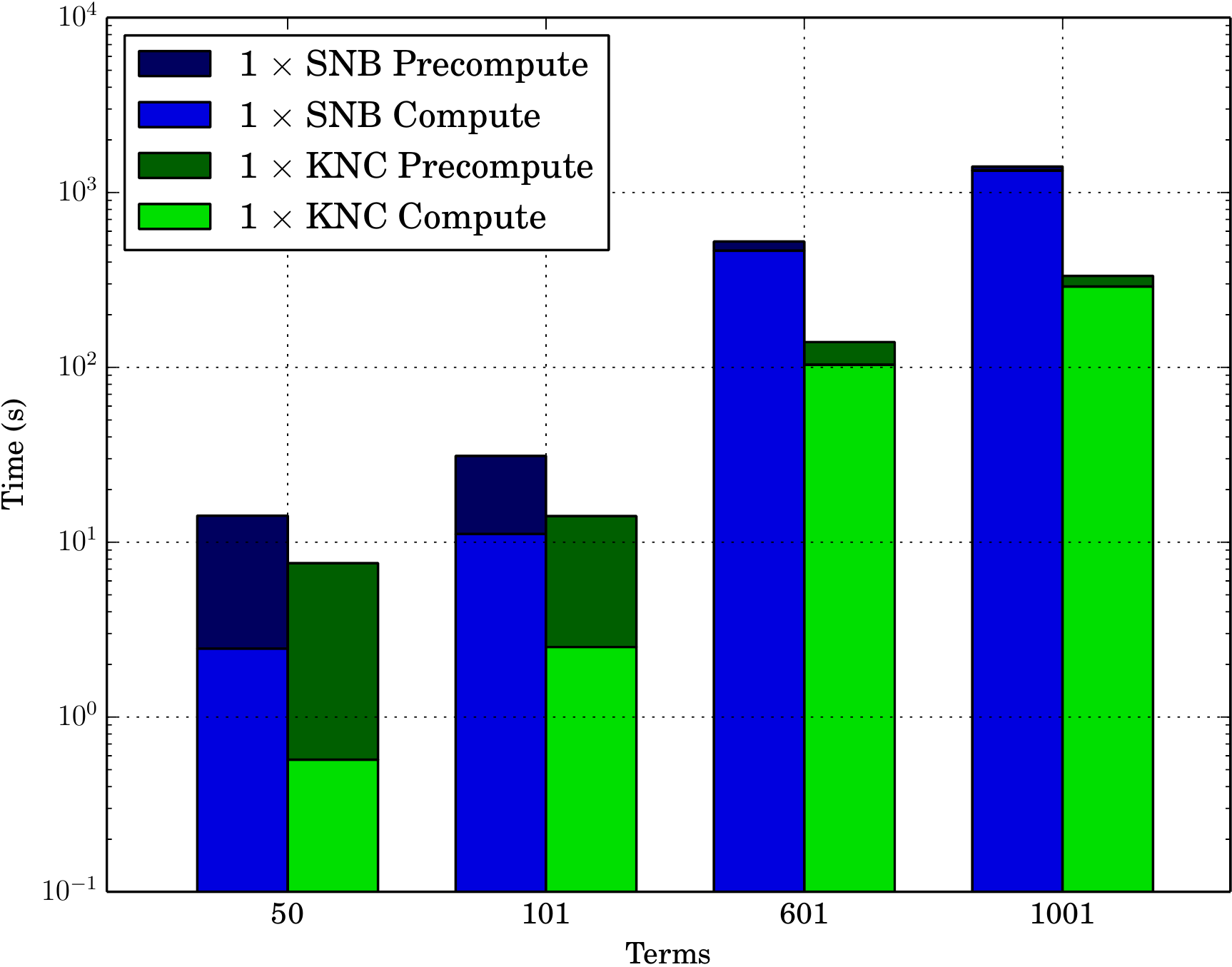}
  \caption{Modal2D}
  \label{fig:problem-sizes-2d}
 \end{subfigure} %
 \hfill %
 \begin{subfigure}[b]{0.48\linewidth}
  \includegraphics[width=\textwidth]{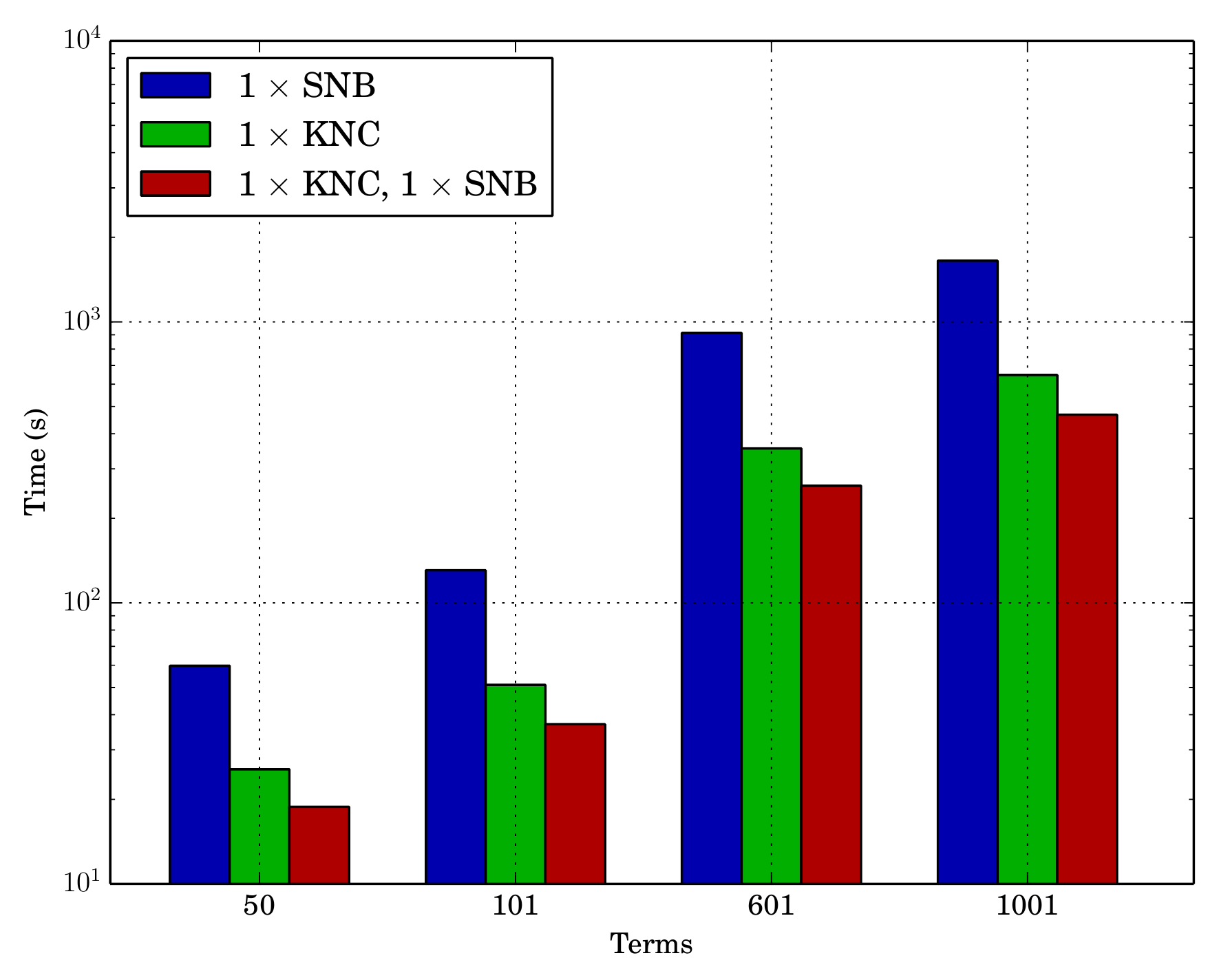}
  \caption{Modal3D}
  \label{fig:problem-sizes-3d}
 \end{subfigure}
 
 \caption{Performance as a function of problem size. In (a) precompute times are stacked on top of the compute times.}
 \label{fig:problem-sizes}
 
\end{figure}

\noindent Figures~\ref{fig:problem-sizes-2d} and \ref{fig:problem-sizes-3d} show how performance scales with problem size for Modal2D and Modal3D, respectively.  As in the previous section, we again report performance for three different configurations: SNB alone; KNC alone; and hybrid execution across both processor and coprocessor. Across all variants and problem sizes, KNC beats SNB. Focusing on the 2D variant, the gap between KNC and SNB is largest for large problems. For the two small problems with 50 and 101 terms, the execution time is dominated by the precompute stage of calculation which does not scale well with increasing numbers of threads, and which is thus relatively expensive on KNC. For the larger problems like the 601 and 1001, the execution time is dominated by the real computation of $\Gamma_{mn}$, which is highly scalable on KNC and thus KNC gains and even larger speed-up of 4.1\x over the SNB for the 1001 case.

For the 3D variant, the gap in execution times between SNB and KNC is more consistent, owing to the large amount of parallelism present even in small problems as a result of our fine-grained decomposition of the loops. All the times in \ref{fig:problem-sizes-3d} are for the same $\ell_1 \ell_2 \ell_3$ space of 18712695 iterations, but for a different number of terms. The amount of computation required in each of Modal3D's functions scales at different rates with respect to the problem size, and this is reflected in the execution times.  The number of eigenmodes that must be integrated scales linearly, but the size of the $Gamma$ matrix and thus the overhead of reducing it scales quadratically.  For all the problem sizes we have tested, integration remains the top hotspot, which is why the scaling is \textit{close} to what one would expect, but the increasing cost of other functions is noticeable.

%% file: conclusions.tex
\section{Conclusions}\label{sec:conclusions}

\noindent We have presented an optimisation study for the 2D and 3D variants of ``Modal'', an early universe simulation and analysis code.  It is representative of two common computational challenges: evaluation of a multi-dimensional integral/sum both on a rectangular dense domain, and on a domain which is neither. The optimisation steps detailed here would be applicable to any similar code.

Through a combination of algorithmic improvements, the introduction of thread-level parallelism, and exposing opportunities for hardware vectorisation, we have achieved significant whole application speed-ups: 1765\x on KNC and 833\x on 2\x SNB in the 2D case; and 108\x on KNC and 83.9\x on 2\x SNB in the 3D case.  In both cases, the greatest source of speed-up is algorithmic change, and the increased amount of exploitable parallelism it brings.  Although still significant, hardware-specific tuning of the new algorithms yields less than 10\x improvement in performance.

Our use of standard programming languages ensures that code changes benefit not only the new \XeonPhi coprocessors, but also \Xeon processors -- and performance improvements are expected to persist on the next generation of processors and coprocessors (codenamed ``Knights Landing'').  Investing in code optimisation and modernisation today can deliver significantly greater gains than waiting for future advances in processor technology and will ensure that we maximise the science done on any given architecture. 

%% file: acknowledgements.tex
\section*{Acknowledgements}
\noindent The authors would like to thank Jeongnim Kim, Larry Meadows and Jason Sewall of Intel Corporation for their insight and assistance in implementing nested parallelism efficiently and Michele Liguori for his part in developing the Modal method.\smallskip

\noindent This research is supported by an STFC consolidated grant ST/L000636/1, and funded in part by the \Intel Parallel Computing Centre program. This work was undertaken on the COSMOS Shared Memory system at DAMTP, University of Cambridge operated on behalf of the STFC DiRAC HPC Facility. This equipment is funded by BIS National E-infrastructure capital grant ST/J005673/1 and STFC grants ST/H008586/1, ST/K00333X/1.

\noindent {\it Disclaimers:} Software and workloads used in performance tests may have been optimized for performance only on Intel microprocessors.  Performance tests, such as SYSmark and MobileMark, are measured using specific computer systems, components, software, operations and functions. Any change to any of those factors may cause the results to vary. You should consult other information and performance tests to assist you in fully evaluating your contemplated purchases, including the performance of that product when combined with other products.  Configurations: see Table~\ref{table:hardware}.  Tests performed by DAMTP, University of Cambridge. For more complete information visit \url{http://www.intel.com/performance}. \smallskip

\noindent Optimization Notice: Intel's compilers may or may not optimize to the same degree for non-Intel microprocessors for optimizations that are not unique to Intel microprocessors. These optimizations include SSE2, SSE3, and SSSE3 instruction sets and other optimizations. Intel does not guarantee the availability, functionality, or effectiveness of any optimization on microprocessors not manufactured by Intel. Microprocessor-dependent optimizations in this product are intended for use with Intel microprocessors. Certain optimizations not specific to Intel microarchitecture are reserved for Intel microprocessors. Please refer to the applicable product User and Reference Guides for more information regarding the specific instruction sets covered by this notice.  Notice Revision \#20110804 \smallskip

\noindent Intel, Xeon, Xeon Phi and VTune are trademarks of Intel Corporation in the U.S. and/or other countries.  Other names and brands may be claimed as the property of others.  

%% file: appendix.tex
\section{Physics of the Modal code}\label{appx:A}

\subsection{Theory}
We observe the temperature $T$ of the cosmic microwave background (CMB) on a distant sphere and we can represent anisotropies in this temperature $\Delta T/T$ as 
\begin{align}
\frac{\Delta T}{T}(\un) = \sum_{lm} a_{lm} Y_{lm}(\un).
\end{align}
where $Y_{lm}(\un)$ are the usual spherical harmonics with multipoles $\ell, m$ with $-\ell \le m\le \ell$. Most quantitative cosmology has developed using the two-point correlation or power spectrum $\langle a_{\ell m}a^*_{\ell m}\rangle$ defined from a average over the azimuthal multipole $m$
\begin{align}
C_\ell = \frac{1}{2\ell+1} \sum_m |a_{\ell m}|^2
\end{align}
However, we are interested in new information from the three-point correlator or bispectrum, averaged over orientations as the CMB is assumed isotropic,
\begin{align}
B_{\ell_1 \ell_2 \ell_3} = \sum_{m_i} \(\begin{array}{ccc}\ell_1 & \ell_2 & \ell_3 \\ m_1 & m_2 & m_3 \end{array}  \) \langle a_{\ell_1 m_1}a_{\ell_2 m_2}a_{\ell_3 m_3}\rangle\,.
\end{align}
This is the spherical respresentation of the 2D CMB bispectrum of triangles on the sky. One of our key goals is to connect this to the 3D primordial bispectrum $\bar{B}(k_1,k_2,k_3)$ generated in the early universe; here, $\bar{B}(k_1,k_2,k_3)$ is defined in terms of wavenumbers $k_i$ and must be projected forward using transfer functions describing the evolution of the Universe to predict the late-time $B_{\ell_1 \ell_2 \ell_3}$.

 The modal method is designed to constrain the bispectrum. If you integrate the bispectrum you get the skewness (\textit{i.e.} how much the distribution leans to one side) of a distribution and, since the bispectrum is zero for a Gaussian distribution, this an effective test of non-Gaussianity.  The issue is that the bispectrum is a full three dimensional quantity and so calculating and measuring it is nontrivial. If we wish to constrain a theory which predicts a bispectrum $\bar{B}(k_1,k_2,k_3)$ at the end of inflation, we first need to evolve it forward to today via convolution with  transfer functions $\Delta$:
\begin{align}
B_{\ell_1 \ell_2 \ell_3} = h_{\ell_1 \ell_2 \ell_3} \int \bar{B}(k_1,k_2,k_3) \int r^2 \(\Pi_{i=1}^3 \, \Delta_{\ell_i}(k_i)  j_{\ell_i}(rk_i)\) dr dk_1 dk_2 dk_3 \,.
\end{align}
where $h$ is a geometric factor related to the projection onto a 2-sphere which will be defined below. This is a 7-dimensional calculation and is impossible in practice. The modal method simplifies this by decomposing the inflationary bispectrum into a set of specially chosen basis functions $\bar{Q}_n$ for which this calculation can be dramatically simplified. These $\bar{Q}_n$ are defined as:
\begin{align}
\nonumber \bar{Q}_n(\klist) &= \frac{1}{6} \( \bar{q}_i(k_1)\bar{q}_j(k_2)\bar{q}_k(k_3) + \bar{q}_j(k_1)\bar{q}_k(k_2)\bar{q}_i(k_3) + \bar{q}_k(k_1)\bar{q}_i(k_2)\bar{q}_j(k_3) \right. \\ &+ \left. \bar{q}_k(k_1)\bar{q}_j(k_2)\bar{q}_i(k_3) + \bar{q}_j(k_1)\bar{q}_i(k_2)\bar{q}_k(k_3) + \bar{q}_i(k_1)\bar{q}_k(k_2)\bar{q}_j(k_3)\)
\end{align} 
where the relation between $n$ to $ijk$ is defined via a pre-determined one-to-one mapping which is optimised for convergence. The mapping both is non-analytic and sparse (so not all, or even most, $ijk$ triples correspond to a n) and is read in from a pre-calculated list. An example mapping would look like:
\begin{align}
\begin{array}{lclll}
n & \rightarrow & i & j & k\\
\hline
0 & \rightarrow & 0 & 0 & 0\\
1 & \rightarrow & 0 & 0 & 1\\
2 & \rightarrow & 0 & 1 & 1\\
3 & \rightarrow & 1 & 1 & 1\\
4 & \rightarrow & 0 & 0 & 2\\
5 & \rightarrow & 0 & 1 & 2\\
\cdots
\end{array}
\end{align}
With this basis we then have $\bar{B}' = (k_1 k_2 k_3)^2 \bar{B} = \sum_n \bar{\a}_n \bar{Q}_n$ where the $\bar{B}'$ is the signal to noise weighted version of $\bar{B}$. The particular form of the basis function allows the projection to be calculated simply and we have:
\begin{align}
\widetilde{Q}_{n\, \ell_1 \ell_2 \ell_3} &\equiv \frac{1}{6} \int r^2dr \(\tilde{q}_{i}(r,\ell_1) \tilde{q}_{j}(r,\ell_2) \tilde{q}_{k}(r,\ell_3) + 5\,\mbox{permutations}\) \\
\tilde{q}_{i}(x,\ell) &\equiv  \int dk k^2 \bar{q}_i(k) \Delta_\ell(k) \, j_\ell(kr)
\end{align}
and now $B_{\ell_1 \ell_2 \ell_3} = \sum_n \bar{\a}_n \widetilde{Q}_n$ and the convolution is effectively 2-dimensional.  However this form still proves difficult to use for estimation because of the radial integral $r$ (\textit{i.e.} distance from inflation to now, along a line of sight).  It is more efficient to use a second basis at the time of observation:
\begin{align}
\nonumber Q_{n\,\ell_1 \ell_2 \ell_3} &= \frac{1}{6} \( q_i(\ell_1)q_j(\ell_2)q_k(\ell_3) + q_j(\ell_1)q_k(\ell_2)q_i(\ell_3) + q_k(\ell_1)q_i(\ell_2)q_j(\ell_3) \right. \\ &+ \left. q_k(\ell_1)q_j(\ell_2)q_i(\ell_3) + q_j(\ell_1)q_i(\ell_2)q_k(\ell_3) + q_i(\ell_1)q_k(\ell_2)q_j(\ell_3)\)\,,
\end{align} 
and to project a signal to noise weighted version of the $\widetilde{Q}$ into this new basis, so that we have 
\begin{align}
Q_n = \sum_m \Gamma_{nm} \widetilde{Q}'_m = \sum_m \Gamma_{nm} \frac{v_1 v_2 v_3}{\sqrt{C_{\ell_1} C_{\ell_2} C_{\ell_3}}}\widetilde{Q}_m
\end{align}
where the $v_i = (2\ell +1)^{1/6}$. If we define the inner product (which is designed to mimic the signal to noise structure of the estimator) as:
\begin{align}
\<A,\, B\>_l &\equiv \sum_{\ell_i} \ \left( \frac{h_{\ell_1 \ell_2 \ell_3}}{v_{\ell_1}v_{\ell_2}v_{\ell_3}}\right)^2 \,A_{\ell_1 \ell_2 \ell_3} \, B_{\ell_1 \ell_2 \ell_3}\,,
\end{align}
where $h$ the previously mentioned geometric factor defined by
\begin{align}
h^2_{\ell_1 \ell_2 \ell_3} = \frac{(2\ell_1+1)(2\ell_2+1)(2\ell_3+1)}{4\pi} \(\begin{array}{ccc}\ell_1 & \ell_2 & \ell_3 \\ 0 & 0 & 0 \end{array}  \)^2\,.
\end{align}
Using this we can then define $\Gamma$ in terms of the inner product as:
\begin{align}
\Gamma_{nm} = \sum_{r} \<Q_n, Q_r \>^{-1} \<Q_r, \widetilde{Q}_m \> 
\end{align}
Thus for optimising the calculation of $\Gamma$ we only need to focus on optimising the evaluation of the inner product.  The majority of the calculation time is in evaluation the second inner product due to the radial integral inherent in $\widetilde{Q}$, so we will restrict our attention to that part alone defining
\begin{align}
\Gamma'_{nm} = \<Q_r, \widetilde{Q}_m \> 
\end{align}
which is Equation~\ref{eq:3D}. The reader is referred to \cite{fergusson2007} and \cite{fergusson2010} for a full explanation of the physics.